\documentclass[12pt]{JHEP3}
\usepackage{epsfig}

\def\gsim{\;\raise0.3ex\hbox{$>$\kern-0.75em\raise-1.1ex\hbox{$\sim$}}\;}
\def\lsim{\;\raise0.3ex\hbox{$<$\kern-0.75em\raise-1.1ex\hbox{$\sim$}}\;}

\def\ignore#1{}  
\def\roughly#1{\mathrel{\raise.3ex\hbox{$#1$%
\kern-.75em\lower1ex\hbox{$\sim$}}}}
\def\lsim{\roughly<}
\def\gsim{\roughly>}

\def\eq{\begin{equation}}
\def\eeq{\end{equation}}
\def\eqa{\begin{eqnarray}}
\def\eeqa{\end{eqnarray}}

\def\wth{3in}

\newcommand{\fourgraphs}[4]{%
\unitlength=1in
\begin{picture}(5.8,4.5)
\put(0,0){\epsfig{file=#3.eps, width=\wth}}
\put(2.9,0){\epsfig{file=#4.eps, width=\wth}}
\put(0,2.3){\epsfig{file=#1.eps, width=\wth}}
\put(2.9,2.3){\epsfig{file=#2.eps, width=\wth}}
\put(0.1,1.9){(c)}
\put(0.1,4.2){(a)}
\put(3.0,1.9){(d)}
\put(3.0,4.2){(b)}
\end{picture}}

\newcommand{\twographs}[2]{%
\unitlength=1in
\begin{picture}(5.8,2.25)
\put(0,0){\epsfig{file=#1.eps, width=\wth}}
\put(2.9,0){\epsfig{file=#2.eps, width=\wth}}
\put(0.1,1.9){(a)}
\put(3.0,1.9){(b)}
\end{picture}}

\newcommand{\bmat}{\left(\begin{array}}
\newcommand{\emat}{\end{array}\right)}

\def\yzero{\smash{\hbox{$y\kern-4pt\raise1pt\hbox{${}^\circ$}$}}}

\def\beq{\begin{equation}}
\def\eeq{\end{equation}}
\def\beqa{\begin{eqnarray}}
\def\eeqa{\end{eqnarray}}

\def\-{\hphantom{-}}

\def\s2{\frac{1}{\sqrt2}}

\def\beq{\begin{equation}}
\def\eeq{\end{equation}}
\def\beqa{\begin{eqnarray}}
\def\eeqa{\end{eqnarray}}

\def\IF{\relax{\rm I\kern-.18em F}}
\def\II{\relax{\rm I\kern-.18em I}}
\def\IP{\relax{\rm I\kern-.18em P}}
\def\IC{\relax\hbox{\kern.25em$\inbar\kern-.3em{\rm C}$}}
\def\IR{\relax{\rm I\kern-.18em R}}

\def\cp{{\cal P}}

\def\Dsl{\,\raise.15ex\hbox{/}\mkern-13.5mu D} 
\def\IZ{Z\kern-.4em  Z}

\def\cp#1{\relax\ifmmode {\IP\kern-2pt{}_{#1}}\else $\IP\kern-2pt{}_{#1}$\fi}

\newcommand{\cO}{{\cal O}}

\newcommand{\bea}{\begin{eqnarray}}
\newcommand{\eea}{\end{eqnarray}}
\newcommand{\be}{\begin{equation}}
\newcommand{\ee}{\end{equation}}
\newcommand{\bt}{\begin{tabular}}
\newcommand{\et}{\end{tabular}}
\newcommand{\ba}{\begin{array}}
\newcommand{\ea}{\end{array}}

\newcommand{\foot}{\footnote}

\newcommand{\mgr}{m_{3/2}}
\newcommand{\agut}{\alpha_{\rm GUT}}

\newcommand{\dg}{^\circ}

\def \soll={\stackrel{!}{=}}
\def \rf=#1{\stackrel{(\ref{#1})}{=}}

\def \dgs{\delta_{\rm GS}}

\def \gev{{\mbox{~GeV}}}
\def \tev{{\rm TeV}}

\def \eg{{\it e.g.\ }}
\def \ie{{\it i.e.\ }}

\title{Selecting Supersymmetric String Scenarios From Sparticle
Spectra}

\author{B.C. Allanach$^{1}$, D. Grellscheid$^{2}$, F. Quevedo$^{2}$ \\
$^{1}$ CERN Theory Division, CH-1211 Geneva 23, Switzerland\\
$^{2}$ DAMTP, CMS, Wilberforce Road, Cambridge CB3 0WA, UK}

\abstract{We approach the following question: if supersymmetry is discovered,
how can we select among different supersymmetric extensions of the
Standard Model? In particular, 
we perform an analysis of the sparticle spectrum in 
low-energy string effective theories, asking which observables best
distinguish various scenarios.  
We examine scenarios differing by the fundamental string scale and
concentrate on GUT and intermediate scale models. 
We scan over four parameters (two goldstino angles, $\tan\beta$ and the
gravitino mass) in each scenario, finding ratios of sparticle
masses that provide the maximum discrimination between them.
The necessary accuracy for discrimination is determined in each
case. We find that 
the required accuracy on various sparticle 
mass ratios is at the few percent level,
a precision that may be achieved in future linear colliders.
We also map out phenomenologically viable regions of parameter space.}

\preprint{CERN-TH/2001-223\\ DAMTP-2001-74\\ hep-ph/0111057}

\begin{document}
\section{Introduction}
There is great expectation in the high energy physics community of the
possibility of discovering evidence of low-energy supersymmetry (SUSY)
during the next few years. 
The smoking-gun signatures of production and detection of super-partners could
be observable at the Tevatron, the Large Hadron Collider and a future linear
collider facility. If such signatures are detected, we may enter a new era of
high energy 
physics with closer contact between fundamental theory and experiment. Since
there are many supersymmetric models,
the issue may then turn from discovering SUSY to selecting and 
eliminating the different supersymmetric models that have been proposed over
 the years. It is therefore useful to 
examine ways of comparing the different models.
Experiments may start to
 give us information not only about particle content but also
about the mechanism of SUSY breaking and (eventually) the
messenger of this breaking.

Many argue that the best motivated models of low energy SUSY are 
those that can be derived from string theory. Even though there is a 
plethora of possible specific string models, we can attempt to perform a 
fairly model independent analysis using some general string scenarios. There
are several ways to parameterize these scenarios. In recent years it was
realised that the underlying scale of a 
fundamental string theory can be different from the Planck scale of
$10^{19}\gev$ if the observable fields are confined to a brane within
a higher dimensional world. Since the size of the extra dimensions
is not necessarily fixed in such a theory, we have the
freedom to argue for different values of the string scale. Two such
possibilities with several indications in their favour are the GUT
scale $M_{GUT}\sim O(10^{16})\gev$~\cite{witten1} and the intermediate 
scale of around $M_I=10^{11}\gev$~\cite{iscale1, iscale2}. 
The GUT scale is favoured by the data in the MSSM-desert gauge unification
picture~\cite{witten1}. The intermediate scale is motivated by a natural
solution to the strong CP problem, the scale of neutrino
masses~\cite{iscale2} and a natural supersymmetry breaking scale in
gravity mediated supersymmetry breaking scenarios.
But in this case, assuming an MSSM-like low energy spectrum, the gauge
couplings evolved from the data at the electroweak scale 
$M_Z$ to $M_I$ do not meet, contrary to the
naive string prediction.
One must address the issue of gauge unification in the
intermediate scale models, and two possibilities have been identified:
one possibility is to achieve precocious
unification \cite{iscale2} through
the inclusion of additional matter fields on top of the MSSM 
content. Explicit intermediate scale string models have sometimes exhibited the
existence of such
superfields extra to the MSSM~\cite{explicitmodels}.
Another possibility is mirage gauge coupling unification~\cite{luis}. In this
picture, gauge unification at the GUT scale is
merely an illusion created by string loop effects; the field theoretic 
gauge couplings are set non-universally at the (intermediate) string scale.
Here, no additional matter fields beyond the MSSM are necessary.

We can imagine having 
experimental data on supersymmetric particles and would like to see if they may
 be able to determine the favoured string scale, or other important parameters
of the fundamental theory. 
In this way we would start to obtain relevant 
information about the fundamental theory at a large scale from measurements at
 low energies. 
Motivated by this possibility we ask to what extent we can 
discriminate scenarios for the 
underlying theory. See \cite{faraggi} for related work.

Supersymmetric phenomenology is extremely complicated. 
The sparticles undergo cascade decays which provide various signatures in
experiments. Such signatures notoriously depend not only on the model
of SUSY breaking, but also upon its parameters. 
If a SUSY breaking parameter changes, mass differences
in the sparticle spectrum can change sign.
Decay channels in the cascade sparticle decays then switch on and off, and the
identified signature is easily lost. 
Such is the complexity of supersymmetric phenomenology that experimental
studies of future colliders have typically focused on a few points in
parameter space~\cite{ATLASTDR}.

To avoid getting bogged down in such complexity and model dependence, and in
order to start the
phenomenology ball rolling, we will assume that some of the superparticles'
masses have been measured. 
We will try to gain information on the high energy
theory from their values.
Explicitly, we will examine three different scenarios:
\begin{itemize}
\item
 String scale at the GUT scale $M_{GUT} \sim O(10^{16})\gev$, defined by 
the scale of electroweak gauge unification $g_1(M_{GUT})=g_2(M_{GUT})$.
\item Intermediate string scale 
($M_I = 10^{11}\gev$) 
with extra leptons to achieve gauge coupling  
unification at $M_I$, which we will refer to as {\it early unification}.
 \item Intermediate string scale ($M_I = 10^{11}$~GeV)  with {\it mirage
unification}~\cite{luis}. The particle spectrum is assumed to be as in the
MSSM, and the gauge couplings at $M_I$ are unequal. 
\end{itemize}
To predict sparticle masses from these scenarios, we must solve the
renormalization group equations
starting from a theoretical boundary condition parameterized by 
the string scale, the goldstino angles, 
$\tan \beta$ and the gravitino mass $m_{3/2}$.
Constraints from experiments and cosmology (if a version of $R$-parity is
conserved, as assumed here) restrict the models.

Previous studies have been performed on intermediate string-type scenarios.
Two of the authors~\cite{aaikq} previously mapped out the spectrum and
naturalness parameter for the dilaton dominated scenario in the GUT, early
unification and mirage scenarios. Baek {\em et al}\/ extended this study to
include bounds from $B\to X_s \gamma$, dark matter and $g-2$ of the
muon~\cite{baek}.
The early unification spectrum and relic neutralino density has also
been investigated~\cite{bkl} when one of the goldstino angles varies.
However, the effects of moduli fields on the SUSY breaking was
neglected in this study. Dark matter observables were also considered 
for mirage or early unification scenarios in ref.~\cite{munoz}.
Dilaton domination (corresponding to a fixed limit of one of the goldstino
angles) is in violation of a charge and colour breaking bound for GUT scale
unification~\cite{munoz2}, but it was found that this bound does not
restrict the intermediate scale case~\cite{aaikq}.

Here we extend previous investigations in three directions. Firstly, we 
scan over both of the goldstino angles.
Secondly, we look for the combinations of sparticle mass measurements which
provide the most discrimination between the string scenarios, once parameters
are scanned over.
Thirdly, we approach the question: is it possible to select a string scenario
from a knowledge of the masses of the different sparticles? We re-phrase this
question as: what accuracy in sparticle mass measurements is required to 
select a string scenario?

\section{Soft SUSY Breaking Terms}
The Type I string models considered here are based on orientifold
compactifications of type IIB strings, and share a number of
similarities with the well studied phenomenological models 
derived from compactified heterotic string theories~\cite{bim}.
Both contain a dilaton superfield $S$, and moduli fields $T_i$
connected with the size and shape of the extra dimensions. These two fields
contribute to SUSY breaking when their auxiliary fields
acquire vacuum expectation values (VEVs). 
We consider one overall modulus $T$ and later one blowing-up 
mode $M$ as a parameterization of the stringy SUSY breaking.
The respective contribution of $F^S$ and $F^T$
 to the SUSY soft breaking terms can be parameterized in a goldstino
angle $\theta$ (where $\sin\theta=1$ corresponds to dilaton
domination and $\cos\theta=1$ denotes moduli domination).
 
One difference between heterotic and type I string models is
that the string scale is not constrained to be near the Planck mass. 
Furthermore, the 
set of moduli fields connected to the blowing-up modes, $M_\alpha$, 
play a more relevant role for SUSY breaking
in Type I models than in the heterotic ones.
They contribute explicitly to the gauge kinetic coupling and therefore 
their $F$-terms may induce gaugino masses and the other soft 
breaking terms. $M$ mixes in the K\"ahler potential with the
 modulus field $T$. Accordingly it is
convenient to introduce a second, $(F^T,F^M)$ mixing angle $\phi$ 
in analogy to the usual dilaton-moduli mixing angle $\theta$. 
To summarise,
\beq
\bmat{c}F^S\\F^T\\F^M\emat=\bmat{c}\sin\theta\\\cos\theta\sin\phi\\\cos\theta\cos\phi\emat
F_{total},
\eeq
where $F_{total}=\sqrt{F_S^2+F_T^2+F_M^2}$.

The introduction of $\theta$ and $\phi$ provides a 
convenient way to parameterize
the influence which the various VEVs have on the soft terms. 
We will generalise the results of ref.~\cite{aaikq} where
only the dilaton domination case was considered
extensively. Here, we examine a wider $\theta$ and $\phi$
range but leave out $\theta <10\dg$.
For such small values of $\theta$, the string-induced high
scale 
boundary conditions 
on the 
soft breaking terms 
 reach a similar magnitude as the one-loop order anomaly mediated SUSY breaking
terms, which therefore cannot be neglected. The complete set of soft
breaking terms of combined anomaly and gravity mediation has not yet been
computed~\cite{baggerandpoppitz}. We therefore leave this slice of parameter
space to the future, when such terms might be included in an analysis.
Like the analysis in~\cite{aaikq}, we assume that the full standard model
gauge group 
arises from a single brane and that SUSY breaking is dominated by the
$F$ terms of the $S$, $T$ and $M$ fields. 
This results in the following soft breaking terms in the
$T+\bar{T}\to\infty$ limit: 

\subsection{Scalar masses}
All scalar masses receive the universal soft SUSY breaking
term
\be\label{m0}
m_0^2=V_0+\mgr^2\left(1-\frac{3}{k}
           C^2\cos^2\theta\sin^2\phi\right)
           +\cO\left({1\over(T+\bar T)^2}\right)\,, \label{mmm}
\ee
where $C=\sqrt{1+{V_0\over3\mgr^2}}$, and $V_0$ is the vacuum
energy. $k$ depends on the form
 of the K\"ahler potential and can be set constant\footnote{Usually 
in heterotic models the scalar masses are
non-universal, since the coefficient of $\cos^2\theta$ depends on the
so-called modular weights of the corresponding matter
fields~\cite{cvetic,bim}. If we considered several $T$ fields here, we
could also expect a non-universal behaviour, parameterized by
different values of
the constant $k$. However, for a single $T$ field we get the universal
behaviour
shown in (\ref{m0}), which amounts to saying that in
these models  all matter fields have modular weight $-1$. For a recent
discussion of non-universal  soft terms in a different context, see for
instance~\cite{xerxes}.}~\cite{aaikq}.
We set $k=3$ in this article, which avoids negative scalar mass squared values
at the string scale. Higher values of $k$ are possible, this would
lead to a weaker dependence of the scalar soft masses on the goldstino angles. 
As usual we will take $C=1$ corresponding to $V_0=0$, \ie a vanishing
cosmological constant.

\subsection{Gaugino masses}
The gaugino masses are equal to
\be\label{m12}
 M_a=\sqrt3C\mgr{\alpha_a\over\agut}\left(\sin\theta
      -{s_a\over8\pi}\agut\,\cos\theta\left(
            {\dgs\over\sqrt{k}}\sin\phi + \cos\phi\right)
      \right)+\cO\left({1\over(T+\bar T)^2}\right). \label{gag}
\ee
Here $\dgs$ is the Green-Schwarz term coming from anomaly cancellation
(like in compactified heterotic string models). Its value is
a model dependent negative integer of order $\cO(-10)$. We will fix 
its value to $-10$ from here on\footnote{%
Different values for $\dgs$ affect the dependence of $M_a$ on the
goldstino angle $\phi$. While this is negligible for $|\dgs |\lsim
\frac{1}{\agut}$ where the minimum of $M_a$ stays very close to $\theta=0$ on
the $(\theta,\phi)$ plane, higher
values of $\dgs$ lead to a significant perturbation away from
$\theta=0$. For a discussion of variable $\dgs$, see~\cite{manuel}.
}.
Note that, in general, the gaugino masses have a non-universal boundary
condition 
at the string scale. If the string scale is intermediate, 
$\alpha_{a=1,2,3} \neq \alpha_{GUT}$ provides non-universality in
equation~(\ref{gag}) from the first term. The second term provides
non-universality at the string-scale through one-loop stringy effects
away from the dilaton dominated limit ($\theta=90\dg$).
This term is proportional to the model-dependent parameters $s_a$.
In order to make mirage unification possible,
$s_a=2\pi\beta_a$ was chosen~\cite{luis}, 
where $\beta_a$ are the usual MSSM renormalization
 $\beta$-function coefficients $(33/5,1,-3)$.

\subsection{A-terms}
Under the assumption that the Yukawa couplings are moduli-independent,
the trilinear couplings are
\be\label{mA}
A_{\alpha\beta\gamma} = -\sqrt3C
\mgr\bigg(\sin\theta+\cos\theta\cos\phi\hat{K}'
         \bigg)+\cO\left({1\over(T+\bar T)^2}\right)\,.\label{aaa}
\ee
Here $\hat{K}=\hat{K}(M+M^*-\delta_{GS}\log\,(T+T^*))$, the $M$- and $T$-dependent part of the K\"ahler potential.
One can set $\hat{K}^{'}=0$~\cite{aaikq},
 since all fields are assumed to be in the
minimum of the potential, for which the argument of $\hat{K}$ vanishes.

\section{Renormalization Group Analysis}
In order to analyse the sparticle spectra 
 we use the {\tt ISASUGRA} part of the {\tt ISAJET 7.51} package~\cite{isajet} to obtain the
SUSY spectrum starting from the high-energy boundary conditions detailed above.

\subsection{Constraints}
We use the following experimental constraints to 
limit the scenarios~\cite{pdg,susywg}:
\be
m_{\tilde\chi^0_1}>45\gev\qquad 
m_{\tilde\chi_1^{\pm}}>103\gev\qquad 
m_{h_0}>113.5\gev. \label{constr}
\ee

We will also use constraints from the 
recently measured muon anomalous magnetic moment
$a_\mu=(g-2)/2$. The experiment E821 at Brookhaven National Laboratory (BNL)
 reported the measurement~\cite{bnl}\ 
\be
a_\mu\ =\ \left(11659202\pm 14\pm 6\right)\times 10^{-10}.
\ee
The precision in the measurement already significantly constrains 
the SUSY parameter
space and will improve in the future. We will constrain all investigated
models to be within  
90\% CL of the central value of the measurement~\cite{narison}, \ie
\beq
-4.2\times 10^{-10}<\delta a_\mu <41.3 \times 10^{-10}. \label{amunew}
\eeq
The final constraint is the absence of a charged lightest supersymmetric
particle (CLSP). This constraint must be applied when the LSP is stable on
cosmological time scales.
This is the case here because we implicitly assume
that a version of $R$-parity holds, as recently found in specific
string models~\cite{norm}.

\subsection{String scale boundary conditions}
As mentioned above, we consider three overall scenarios: GUT scale
unification (with the string scale set to $M_S=2\times 10^{16}\gev$), early
unification and 
mirage unification (both with $M_S=M_I=1\times 10^{11}\gev$). 
The MSSM spectrum is assumed in each case, except for early
unification where $2 \times L_L + 3 \times E_R$ vector-like
representations are added to the MSSM in order to achieve gauge unification at
$M_I$~\cite{aaikq}. It is assumed that these extra representations have 
negligible Yukawa couplings. We add their effect to the MSSM gauge 
$\beta$-functions above $1\,\tev$. 

The string-scale boundary conditions on the soft terms were obtained from
equations~(\ref{m0}-\ref{mA}), with
$\mgr$, $\theta$ and $\phi$, as well as $\tan\beta$ as free parameters.  
Throughout the whole analysis $\mu>0$ and $m_t=175\gev$ were
assumed~\cite{pdg}. 
Negative $\mu$ leads to a negative $\delta a_\mu$, which is limited from
equation~\ref{amunew} to be small in magnitude. 
This means that, for a given
value of $\tan \beta$, the sparticles must be heavy in order to suppress their
contribution to $\delta a_\mu$ (compare figs.~\ref{maps1} and~\ref{maps-negmu}). 
In this limit, effects of the sign of $\mu$
upon the mass spectrum are suppressed. We can therefore safely ignore the 
$\mu<0$ case because its resulting spectra will be included in our $\mu>0$
results. This also means, though, that the sign of $\mu$ cannot be
determined from these spectra, unless the sparticles are too light to be
compatible with $\mu < 0$ for any $\tan \beta$. 

In the GUT case, $\alpha_a=\agut=1/25$ was used in equation~(\ref{m12}) for
all gauge 
groups $a=1,2,3$. In the mirage unification case, however, the gauge
couplings only {\it appear}\/ to unify at the GUT 
scale.
In reality they are set at the intermediate scale
$M_I=10^{11}$~GeV 
to non-universal values\foot{We note that ref.~\cite{aaikq} assumes gaugino
universality, which leads to different
conclusions for the mirage scenario, especially for the
 charged LSP constraints.}. 
In our analysis,
\beq
\alpha_1(M_I)=\frac{1}{37.6},\qquad\alpha_2(M_I)=\frac{1}{27.0},\qquad
\alpha_3(M_I)=\frac{1}{19.8}
\eeq
is the set of values which remains reasonably stable under iterations of
inserting them into equation (\ref{m12}) and running the RGEs again.

For each of the three scenarios that we consider, we scan 
over the free
parameters $\theta$, $\phi$, $\tan \beta$ and $m_{3/2}$.
$\theta=90\dg$ corresponds
to the dilaton domination case considered in~\cite{aaikq}, 
and the SUSY soft mass terms are independent of $\phi$ in this case. 
As mentioned above, values of $\theta <10\dg$ were 
not considered, since in this case tree level soft masses would vanish
and it would not be possible to neglect the one-loop
effects from anomaly mediated SUSY breaking, which might modify the soft 
terms considerably. 
Table~\ref{input} shows the default ranges and increments of scanned parameters. It
also summarises the other parameters that were kept constant. When
different values were used for a plot, we will mention it explicitly.
\TABULAR[t]{|c|c|c|c|c|c|c|c|c|c|c|c|c|}
{\hline\label{input}
&$\theta$ & $\phi$ & $m_{3/2}$ & $\tan \beta$ & $\mu$ & $m_t$ & $V_0$ & $C$ &
$k$ & $\dgs$ & $M_{GUT}$ & $M_I$\\ \hline
range & 10-90$\dg$& 0-90$\dg$ & 50-1500 & 2-50 & $>0$ & 175 & 0& 1 & 3 & -10
& 2$\times 10^{16}$ &$10^{11}$ \\
\hline
}
{Summary of parameters. The first four parameters are scanned over, and their
range is detailed. For the others, their
value is kept constant except for $\mu$ which is constrained to give the
correct value of $M_Z$. 
All massive parameters ($m_{3/2}, m_t, M_I, M_{GUT}$) are given in units of
GeV.} 

\subsection{Sparticle spectra}
In order to get some feeling for the effect of the goldstino angles, we first
illustrate their effect upon the sparticle spectra. We choose the mirage
unification scenario as a first example. The other scenarios do differ from
these
spectra quantitatively, but the effect of the goldstino angles is similar in
each case. In figure~\ref{spec1}, we show the variation of the mirage
unification spectrum with
$\theta$ for $\tan \beta=30$, $\mgr=300\gev$ and $\phi=60\dg$. As mentioned
above, 
$\theta=90\dg$ corresponds to the dilaton dominated limit. From the figure, we
see that larger $\theta$ increases the splittings between sparticles. 
Larger $\theta$ corresponds to larger gaugino masses in eq.~(\ref{gag}) and
the larger gluino mass raises the other coloured sparticle masses in the
running from $M_I$ to $M_Z$. The weak gauginos also show a milder change from 
the larger values of $M_{1,2}$. Sleptons are largely unaffected by the
variation of $\theta$, demonstrating the fact that the mild increase in 
$m_0$ in eq.~(\ref{mmm}) from increasing $\theta$ is a relatively minor
effect.

The ordering of superparticle masses is largely unaffected by changes in
$\theta$, except for the lightest chargino, which crosses the ${\tilde e}_R,
{\tilde \tau}_2$ and ${\tilde e}_L$ lines\footnote{It is difficult to discern
in the figure that the ${\tilde \tau}_2$ and ${\tilde e}_L$ are
quasi-degenerate.}
and the lightest stop, which crosses the heavier chargino line. Each ordering
can correspond to different cascade decay channels, and might be used to
restrict $\theta$ once these decays are observed. For example, 
when the chargino crosses the ${\tilde e}_R$ line, nothing much changes
because the lightest chargino does not have significant couplings to
right-handed selectrons. However, decays into neutrinos and ${\tilde e}_L$ or
${\tilde \tau}_2$ (which can have a significant left-handed component for
large $\tan \beta$) become viable for $\theta > 70\dg$.
\EPSFIGURE[t]{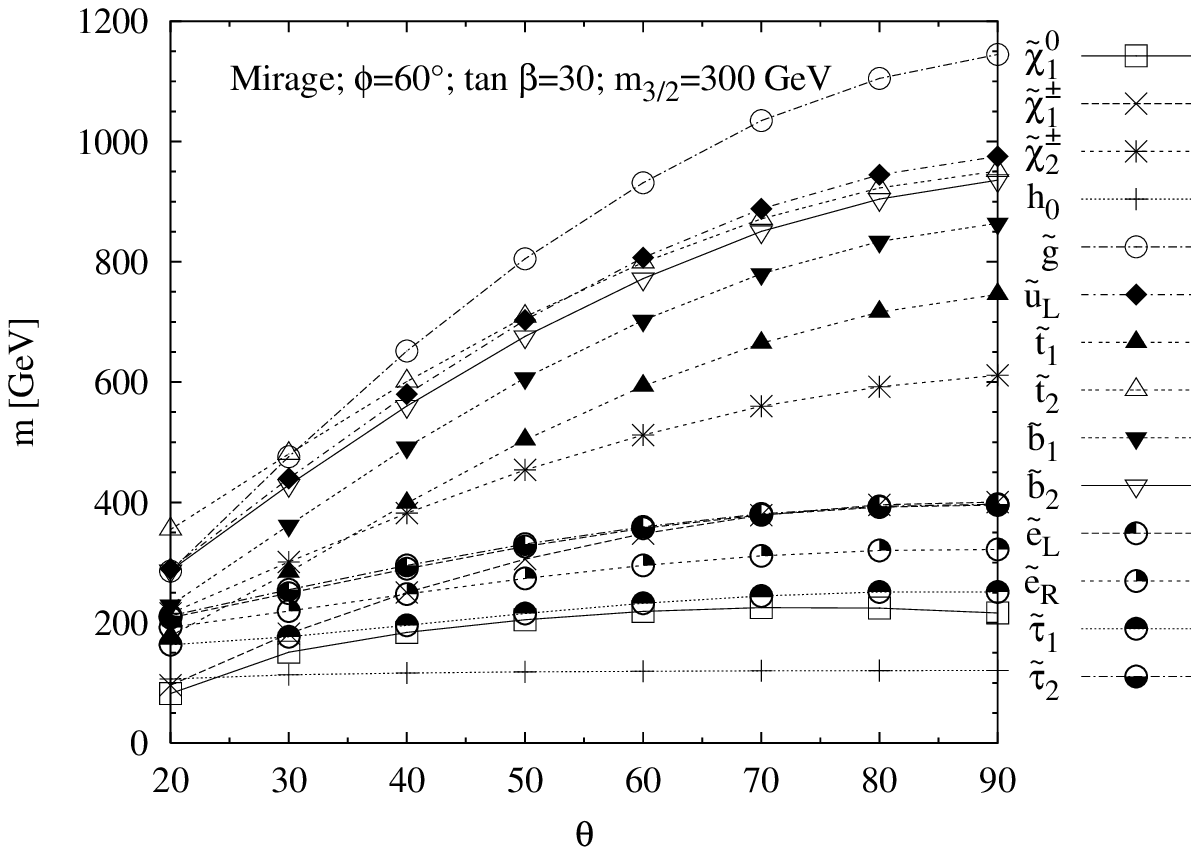}{\label{spec1}Variation of mirage
unification  
sparticle mass spectrum with $\theta$. We use $\phi =60\dg$,
$\tan\beta =30$ and $\mgr =300\gev$. The key on the right-hand side of the
figure details the flavour of sparticle.}
\FIGURE[t]{\label{spec2}
\unitlength=1in
\begin{picture}
(4.5,6.5)
\put(0,0){\epsfig{file=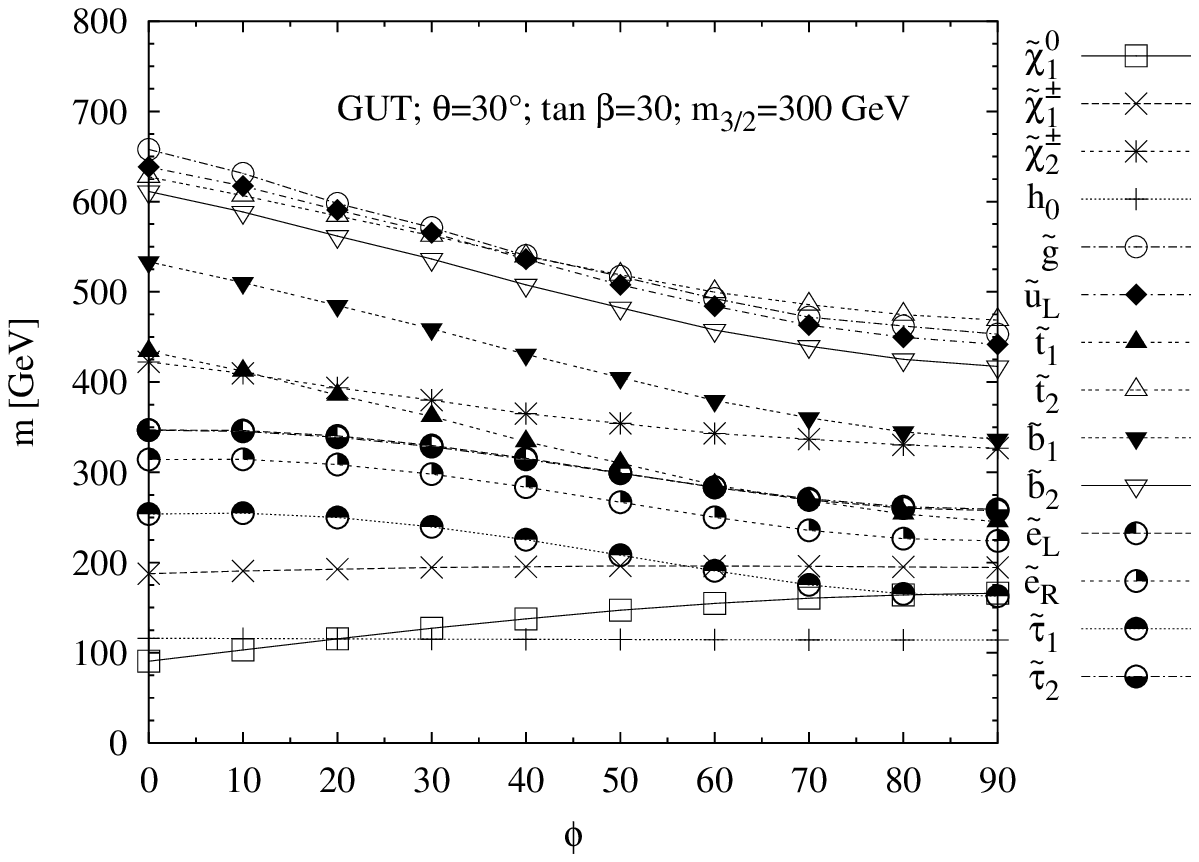, width=4.5in}}
\put(0,3.25){\epsfig{file=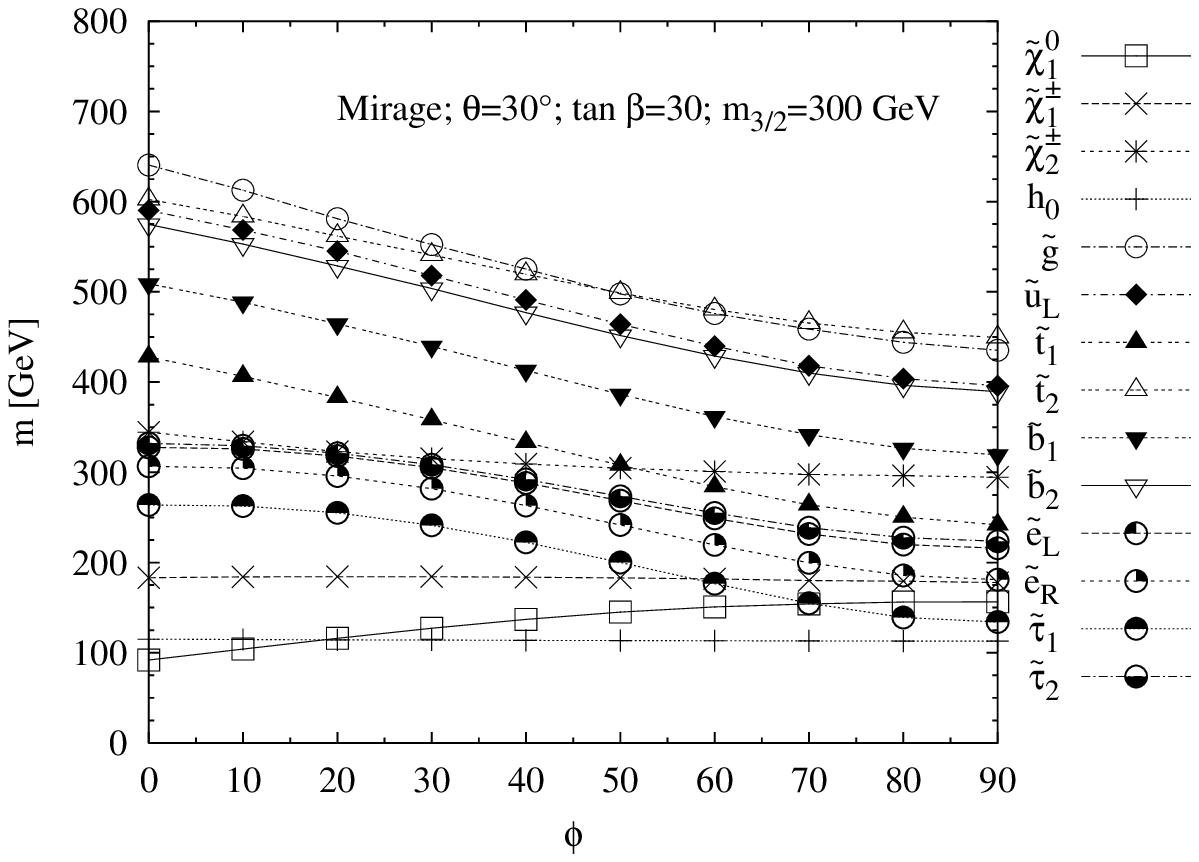, width=4.5in}}
\put(-0.1,6.5){(a)}
\put(-0.1,3){(b)}
\end{picture}
\caption{Variation of sparticle mass spectrum with $\phi$ for (a) mirage and
(b) GUT-scale unification scenarios.
We use $\theta =30\dg$,
$\tan\beta =30$ and $\mgr =300\gev$. The key on the right-hand side of the
figure details the flavour of sparticle.}
}

In figure~\ref{spec2}a, we show the variation of the mirage unification sparticle
spectrum with $\phi$ for $\tan \beta=30$, $\mgr=300\gev$ and $\theta=30\dg$.
Here, increasing $\phi$ has a smaller effect than $\theta$, but generally
decreases the splittings between sparticle masses. The effect is relatively
small because the gaugino masses are only sensitive to $\phi$ through a
suppression factor in eq.~(\ref{gag}). Most of the sparticle masses decrease
with increasing $\phi$, except for the lightest neutralino (which becomes
heavier) and the lightest chargino (which remains roughly
constant). This can be understood by considering the effect of increasing
$\phi$ on the scalar masses in eq.~(\ref{mmm}).
The figure shows that the model would be ruled out for $\phi \geq 70\dg$
because of a stau LSP\@. Aside from this, important changes in the ordering of
sparticle masses occur between the lightest stau and chargino and the heavier
chargino and the lightest stop. The heavier stop becomes heavier than the
gluino for $\phi>50 \dg$. This is potentially important because future
$e^+e^-$ machines 
can produce and measure gluinos more easily through the decays of
pair-produced squarks. In this example, however, the mass difference between
gluino and the heavier stop is not sufficient to produce an on-shell top quark
in a decay. Thus such a process would be highly phase-space suppressed.

Figure~\ref{spec2}b shows the variation of the GUT scale unification sparticle
spectrum with $\phi$ for $\tan \beta=30$, $\mgr=300\gev$ and $\theta=30\dg$.
While quantitative differences exist between figures~\ref{spec2}a and~\ref{spec2}b, it is clear that the two spectra are qualitatively very
similar. There is, however, some re-ordering of sparticle masses between the
GUT-scale and mirage-scale scenarios for certain values of $\phi$. For
example, at $\phi=20\dg$, the heavy chargino and light stop are
interchanged. It would be difficult to discriminate the two scenarios on 
the basis of decays implying a certain ordering of the lightest stop and
chargino since there is no {\em a priori}\/ information about $\phi$, although
it might be possible to fit the masses to certain values of $\theta,\phi$ if
the experimental accuracy were good enough.
For larger values of $\theta$, the variation of the
spectrum with $\phi$ significantly decreases because the terms that depend
upon it in the boundary conditions eqs.~(\ref{mmm})-(\ref{aaa}) become smaller
and sub-dominant to the other $\theta$ dependent terms. 

\FIGURE[t]{%
\fourgraphs{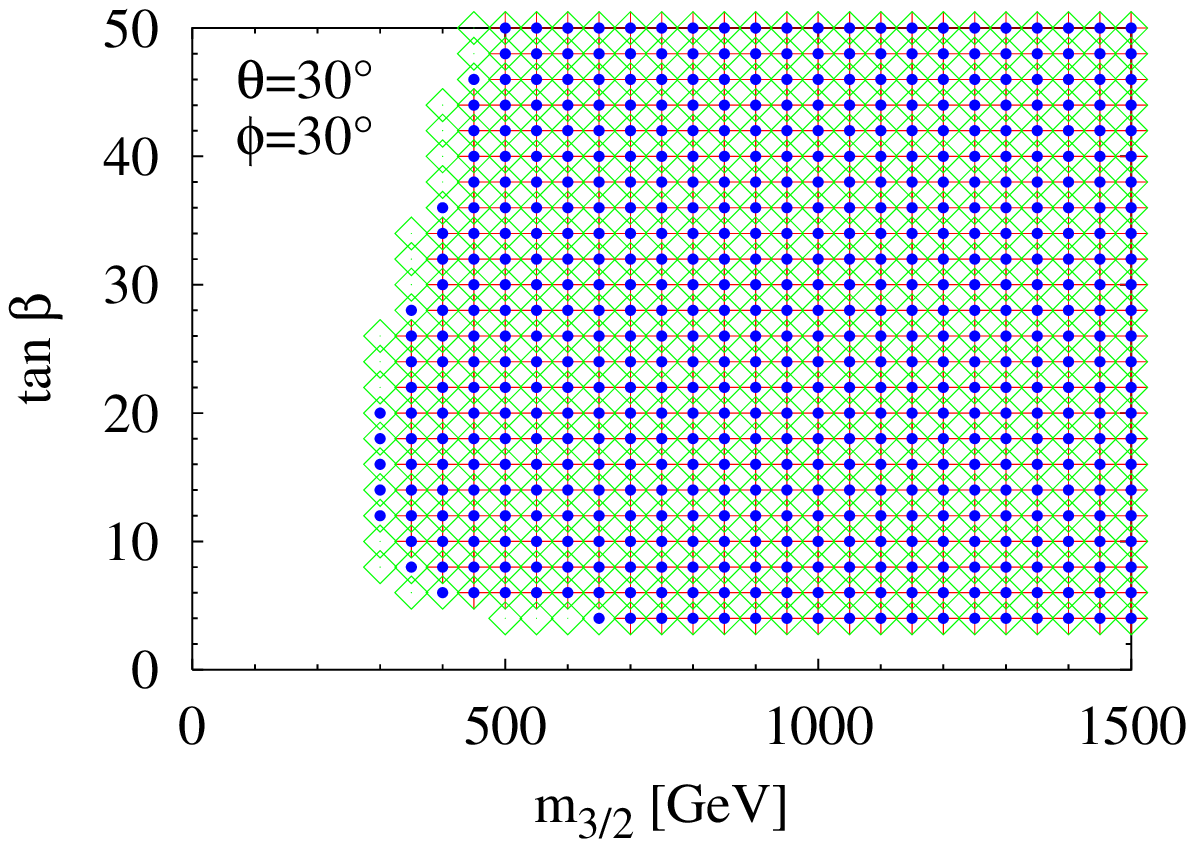}{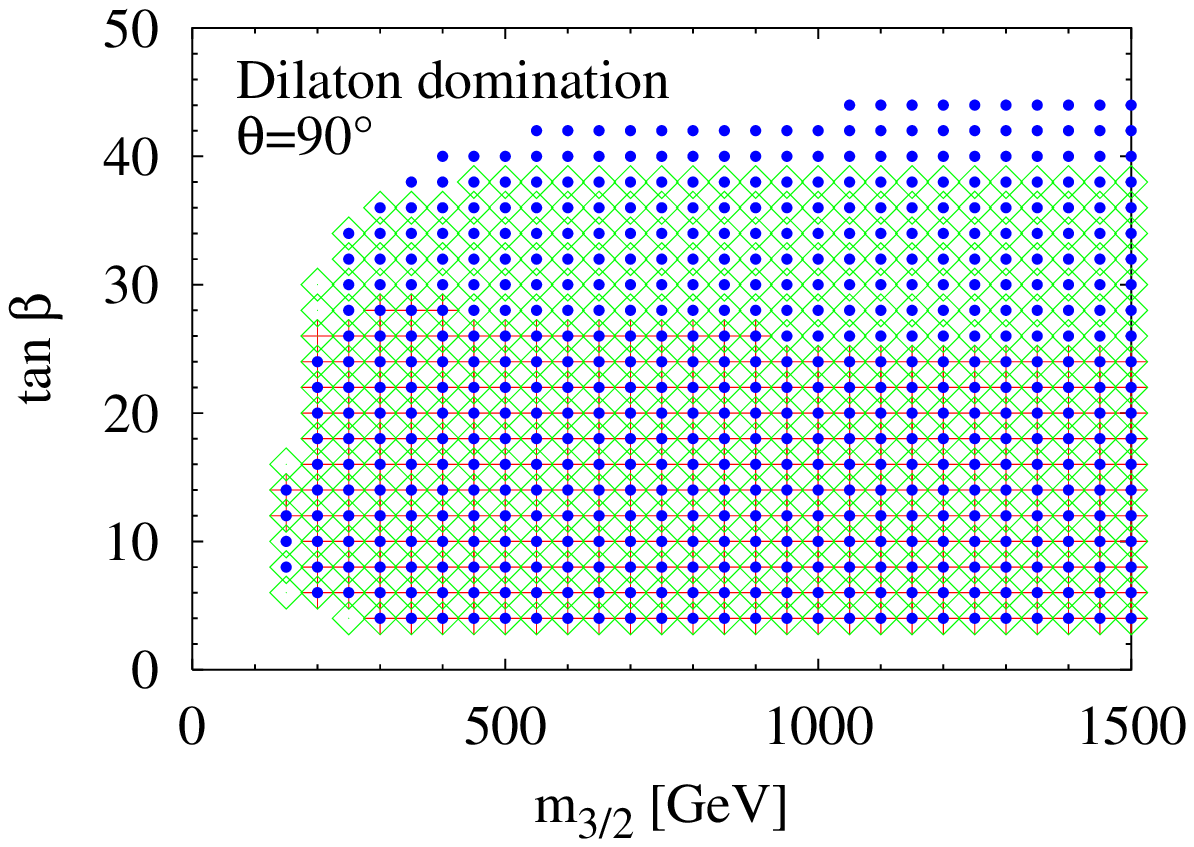}{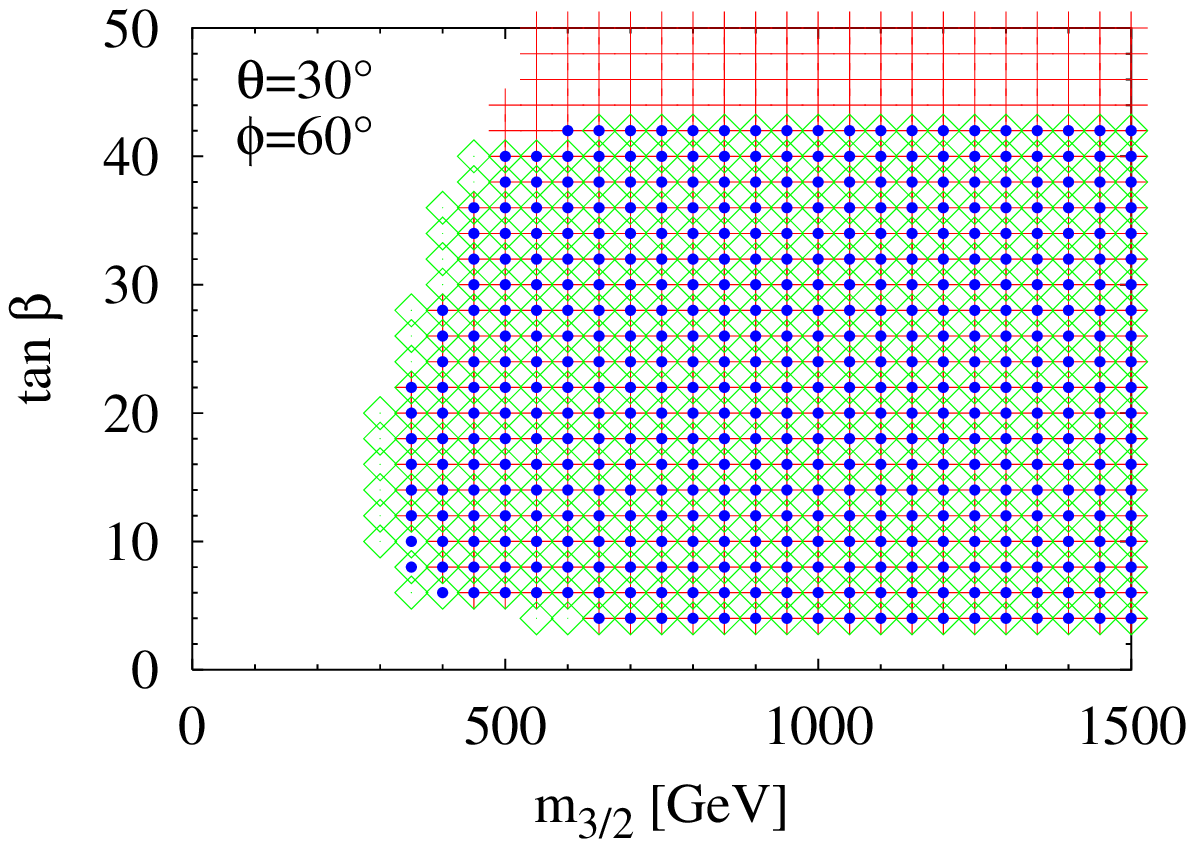}{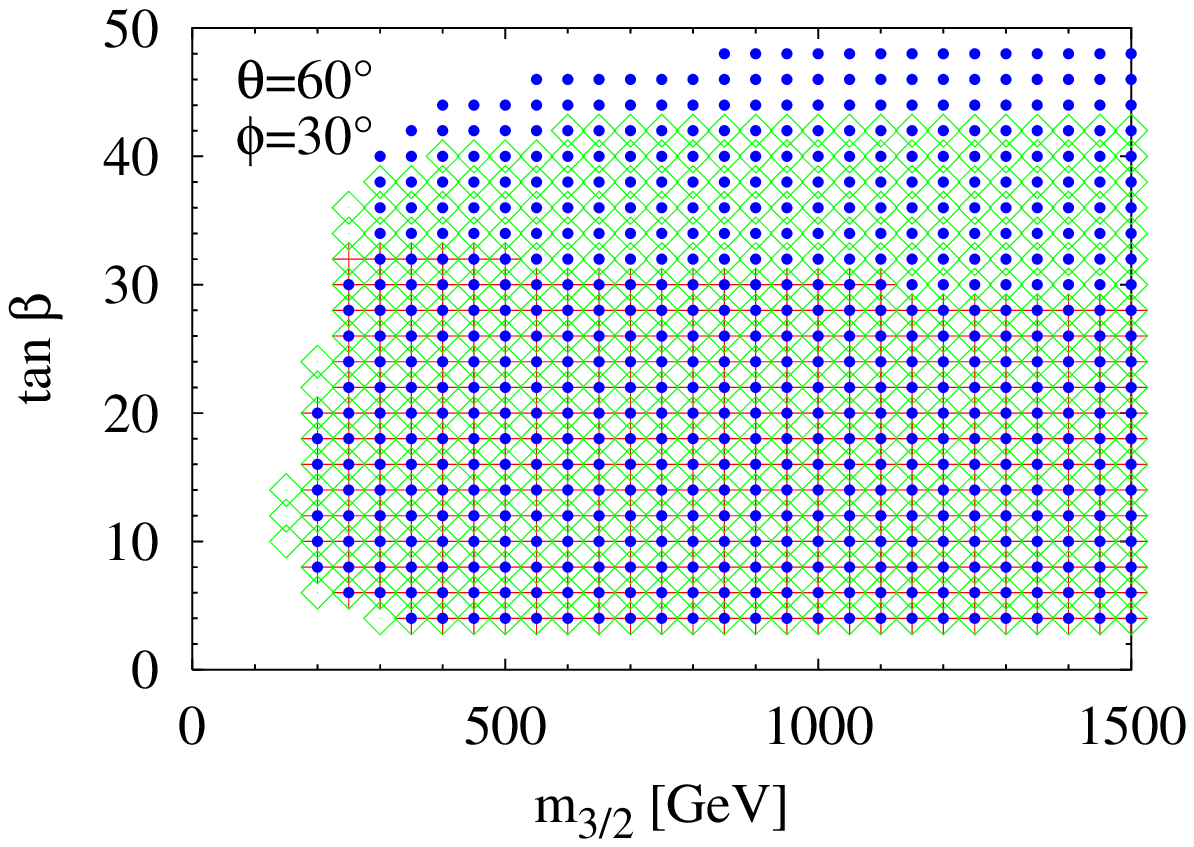}
\label{maps1}
\caption{$\mu > 0$ maps of parameter space for the early unification (red crosses), mirage
(blue dots) and GUT unification scale (green diamonds) scenarios. Here, we
plot viable regions of $\tan \beta$-$m_{3/2}$ parameter space for
fixed values of the goldstino angles, after all exclusion 
limits were applied: 
(a) $\theta=30\dg$, $\phi=30\dg$,
(b) $\theta=90\dg$, $\phi=$arbitrary,
(c) $\theta=30\dg$, $\phi=60\dg$,
(d) $\theta=60\dg$, $\phi=30\dg$. 
}}
\FIGURE[t]{%
\fourgraphs{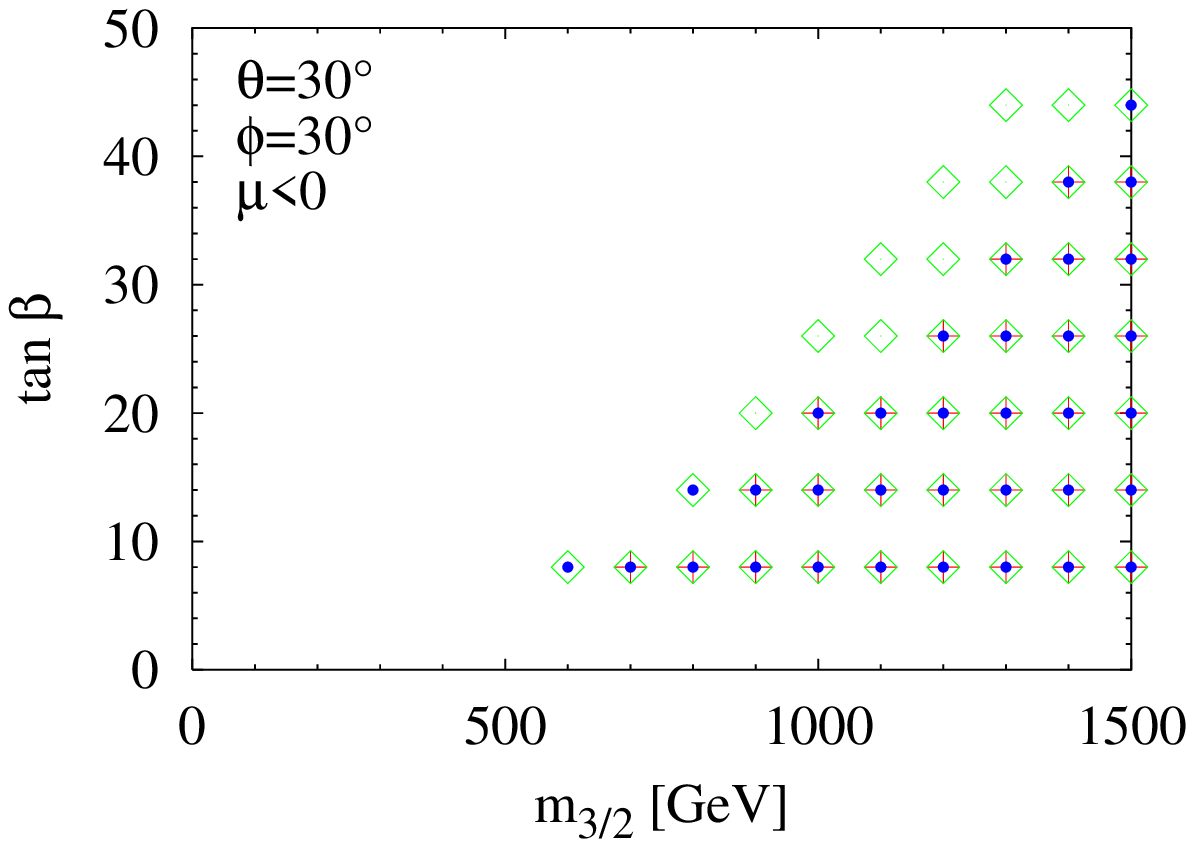}{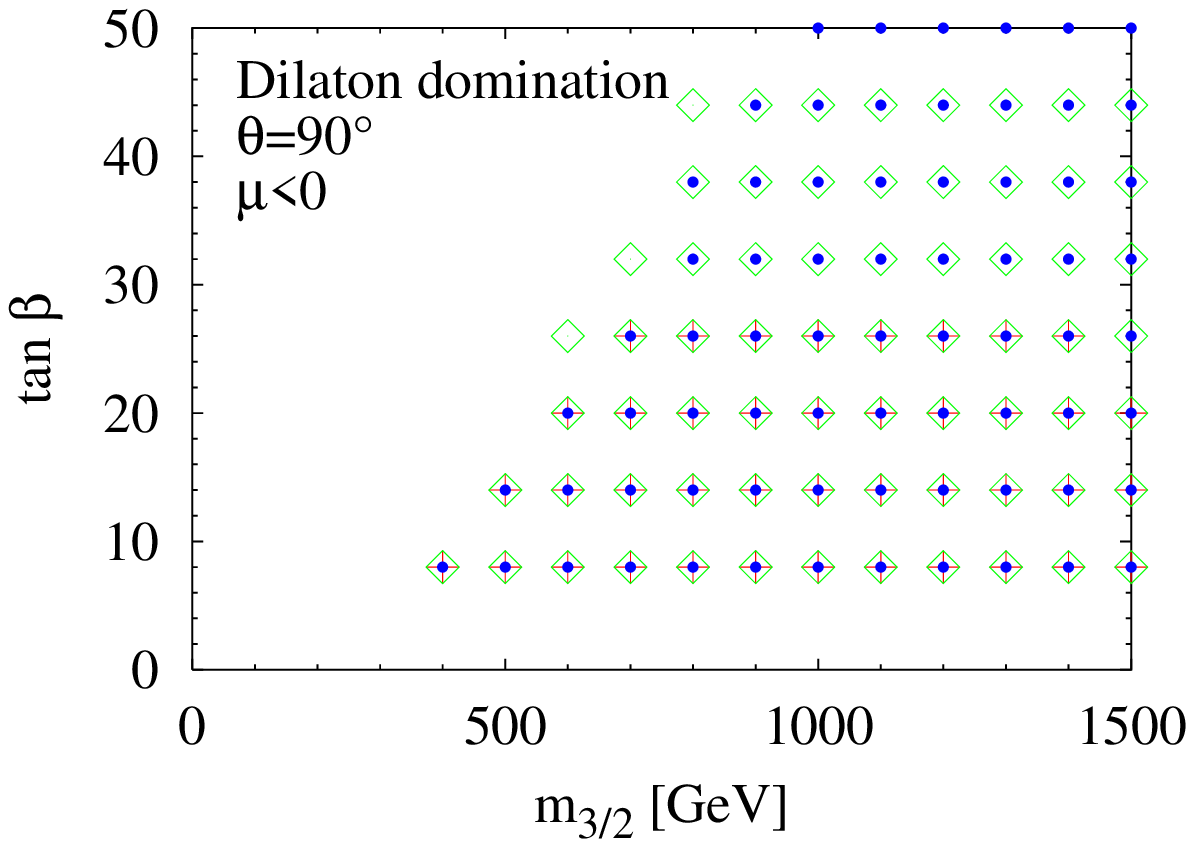}{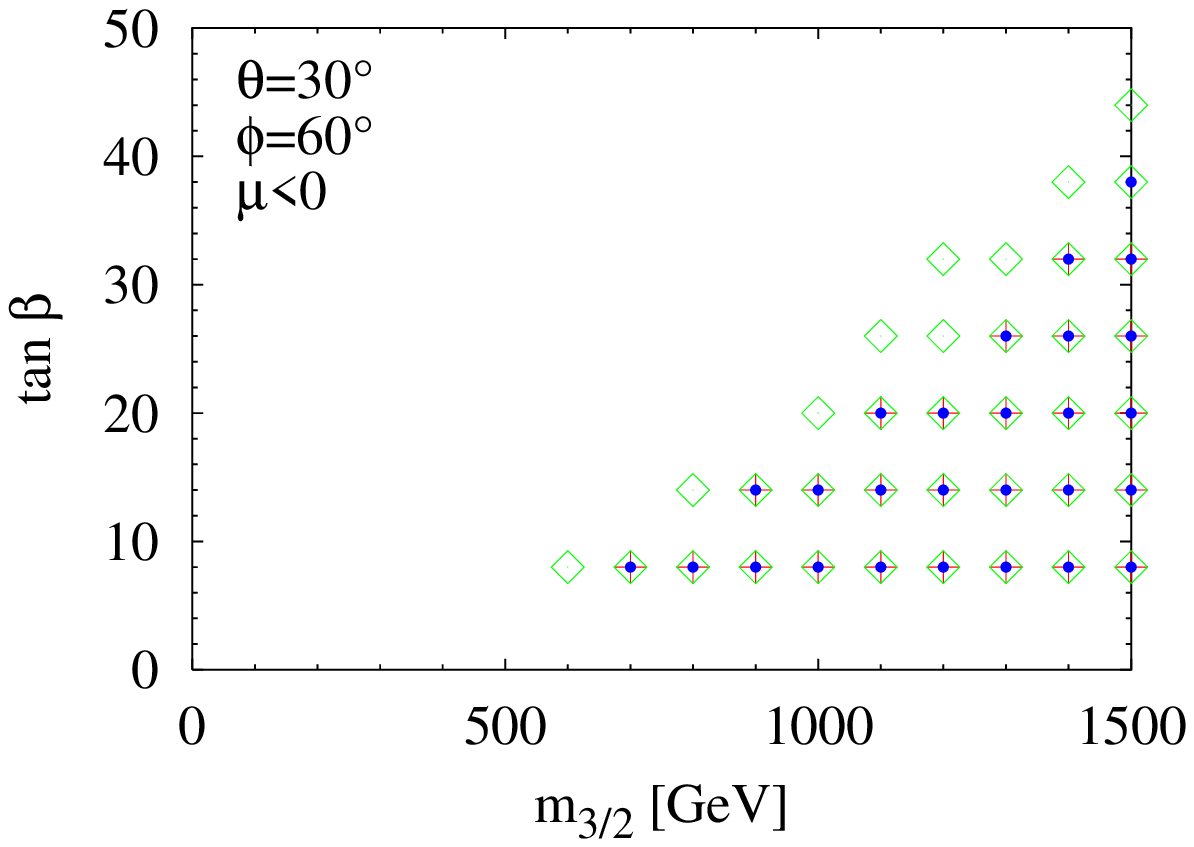}{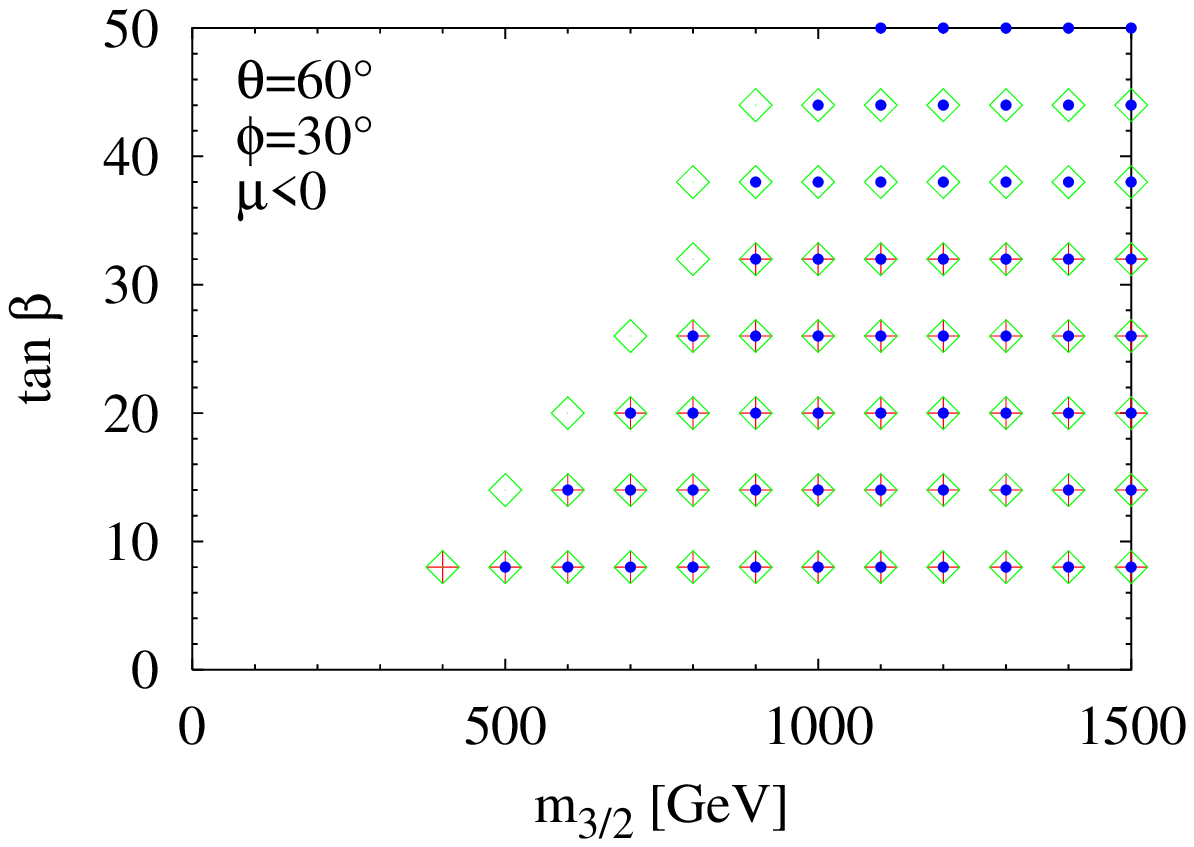}
\label{maps-negmu}
\caption{$\mu < 0$ maps of parameter space for the early unification (red crosses), mirage
(blue dots) and GUT unification scale (green diamonds) scenarios. Here, we
plot viable regions of $\tan \beta$-$m_{3/2}$ parameter space for
for fixed values
of the goldstino angles, after all exclusion limits were applied: 
(a) $\theta=30\dg$, $\phi=30\dg$,
(b) $\theta=90\dg$, $\phi=$arbitrary,
(c) $\theta=30\dg$, $\phi=60\dg$,
(d) $\theta=60\dg$, $\phi=30\dg$. 
}}
\FIGURE[t]{%
\twographs{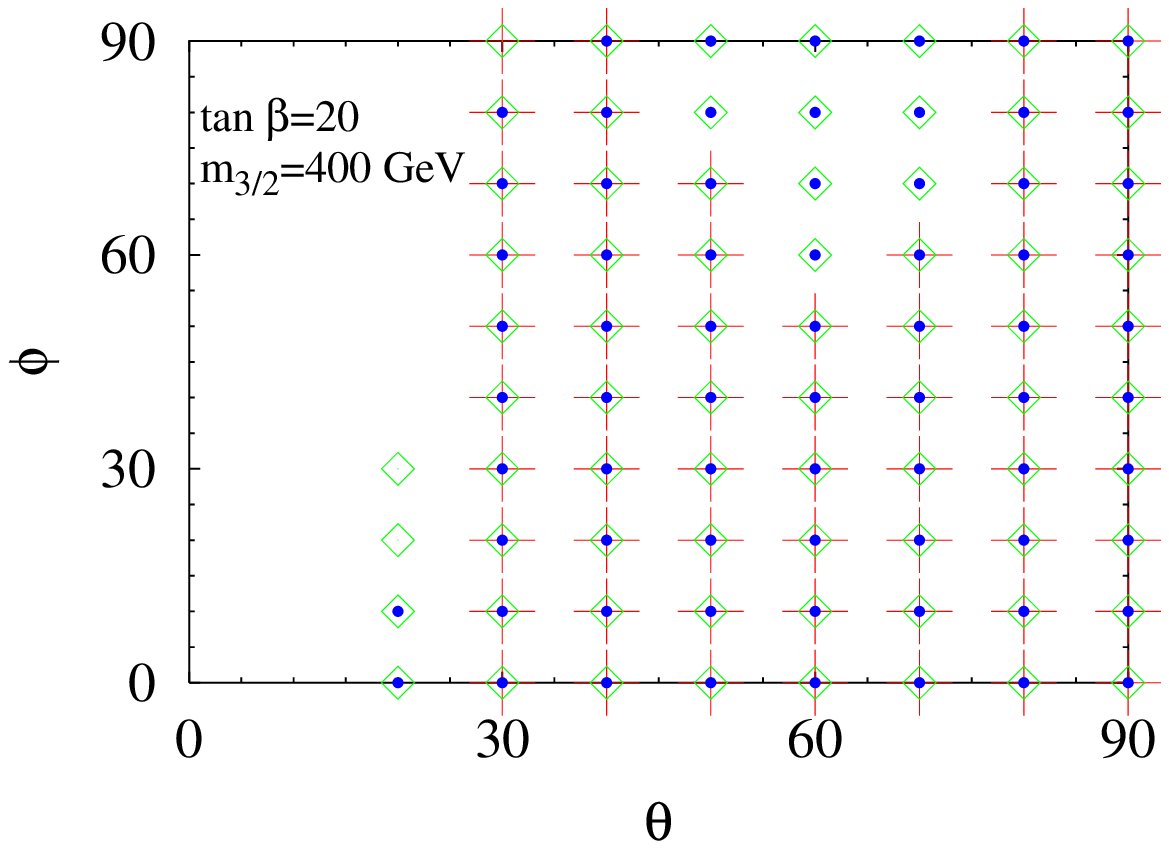}{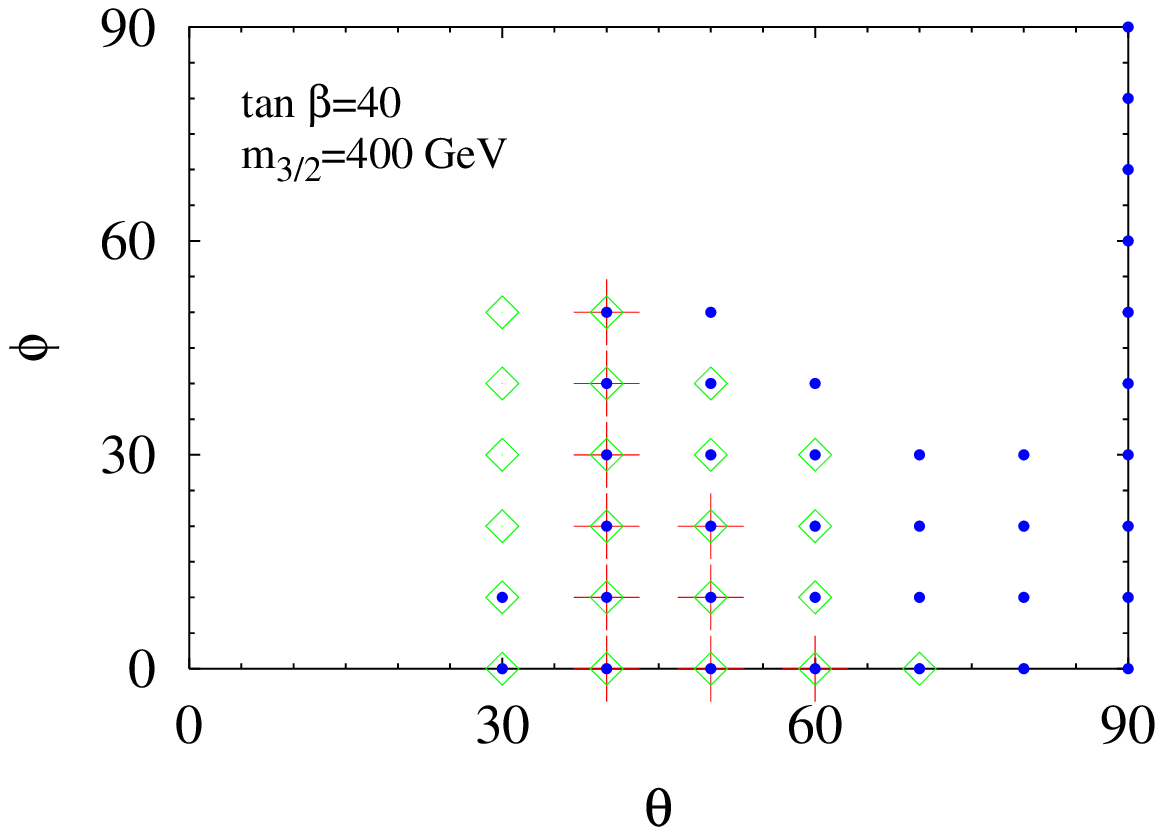}
\label{maps2}
\caption{Maps of parameter space for the early unification (red crosses),
mirage (blue dots) and GUT unification scale (green diamonds) scenarios. Here,
we plot viable regions of $\theta$-$\phi$ parameter space for 
fixed values of $m_{3/2}$ and $\tan \beta$: (a) $m_{3/2}=400$ GeV, $\tan
\beta=20$, (b) $m_{3/2}=400$ GeV, $\tan \beta=40$}
}
\subsection{Maps of parameter space}
We now explore the parameter space, including constraints from the sparticle
mass limits in eq.~(\ref{constr}), requiring that the
$g-2$ anomalous magnetic moment be within 90\% CL of the central measured
value, requiring correct electroweak symmetry breaking 
and imposing a neutral LSP\@.
In figure~\ref{maps1}, we present viable regions of $\tan \beta$-$m_{3/2}$
parameter space for various points in the goldstino angle space and $\mu>0$. 
Dilaton domination has been well covered in the literature~\cite{aaikq, baek}
and so we present other more general regions of the parameter space alongside
the dilaton dominated limit in figure~\ref{maps1}b.
We see from the variation in the viable regions from
figures~\ref{maps1}a-\ref{maps1}d that they are dependent upon the
goldstino angles, which vary between each of the figures. For $\theta=30\dg$
and $\phi=30\dg$, we see from figure~\ref{maps1}a that there is little
dependence of the viable region upon the scenario assumed. The other three
figures show a larger difference between the three scenarios, apart from for 
$\phi=60\dg, \theta=30\dg$, which does not distinguish
the GUT and mirage unification regions significantly, as can be seen
from fig.~\ref{maps1}c. 
In figure~\ref{maps-negmu}, we show the available parameter space for
$\mu<0$. By comparing figs.~\ref{maps1} and~\ref{maps-negmu}, we see that
$\mu<0$ admits less parameter space. This is because the $g-2$ constraint
restricts sparticles to be heavy for $\mu<0$ in order to give a small
contribution to the magnitude of the anomalous moment. The lower bounds upon
sparticle masses are much
stronger for high $\tan \beta$, where the contributions to the anomalous
moment are largest. Figure~\ref{maps-negmu} illustrates that (for low $\tan
\beta$), there is a lower bound $m_{3/2}\gsim 400$ GeV.
 
There are often also bounds upon $\tan \beta$. Low $\tan \beta$ is ruled out
by the direct lower bound upon the Higgs mass from LEP2~\cite{pdg}, whereas
high $\tan \beta$ is either ruled out by the constraint of a neutral LSP, or
of not too {\em large}\/ a supersymmetric contribution to $g-2$ of the muon.
For instance in the dilaton domination $\mu>0$ $(\mu<0)$ case we can see that
the  GUT scale scenario restricts $\tan\beta<40$ (44), the early unification
scenario requires $\tan\beta<30$ (28) and mirage unification constrains $\tan
\beta < 46$. For the lower value of $\theta=30\dg$ in the $\mu>0$ case, the
$\tan \beta$ bound 
disappears for $\phi=30\dg$ in all scenarios as in figure~\ref{maps1}a, but
reappears for higher $\phi=60\dg$ in the early and GUT-scale unification
scenarios (figure~\ref{maps1}d). 

In figures~\ref{maps2}a and~\ref{maps2}b, we show 
the viable regions in goldstino angle 
space for fixed $m_{3/2}=400$ GeV and $\tan \beta=20,40$ respectively.
For lower $\tan \beta=20$,
there are viable dilaton-dominated cases for each scenario.
However, for $\tan \beta=40$ in figure~\ref{maps2}b, dilaton domination
($\theta=90\dg$) is not viable except for the mirage unification case.
The bounds
$\phi \leq 50\dg$ and $\theta<70\dg, 60\dg$ apply
to the GUT-scale and early
unification scenarios respectively. 
For
$\phi=0\dg$, 
there is still a large contribution to unification-scale sparticle masses from
terms involving $\theta$, so a lower bound on $\phi$ does not exist for 
general $\theta$ from sparticle mass bounds. 

\subsection{Discriminating ratios}
We have searched through ratios of masses of sparticles to see which ones
provide the largest discrimination between the three scenarios examined.
The ideal situation would be to find ratios that are predicted to be
completely disjoint regions for the three scenarios. 
Then, if the mass ratios were
determined in an experiment, a comparison with these disjoint regions should
reject or confirm one scenario. In general, this may not be possible so we
resort to finding ratios which discriminate between some of the scenarios in
some cases.
Where disjoint regions are found, we will focus on the accuracy required
in any mass ratio measurement that would be enough to discriminate
two scenarios. 
This accuracy we will define as follows: supposing the central value of a
measurement of a ratio lies on the edge of one of the regions.
The fractional accuracy in each direction that would exclude each of the other
scenarios is taken. Then, we examine all points lying within a region and
quote the lowest fractional accuracy on each ratio required to separate the
regions.
One ratio on its own does not provide enough separation once the
parameters in 
each scenario are scanned over; we therefore resort to considering
combinations of two ratios in order to see if the regions of each scenario are
separated in two dimensional ratio space.

There are a few combinations of ratios that
provide near separation. We display the best two such combinations in
figure~\ref{allscat}. 
For each point in the parameter scan described in table~\ref{input},
that satisfies the constraints in eq.~(\ref{constr}) as well as the neutral
LSP constraint, one point appears on the plot. 
Here, we have taken input parameters {\em at random}\/ within the ranges defined
in table~\ref{input} for 10000 points (as is the case for the rest of this
subsection). 
\FIGURE[t]{%
\label{allscat}
\unitlength=1in
\begin{picture}(6,2.4)
\put(0,0){\epsfig{file=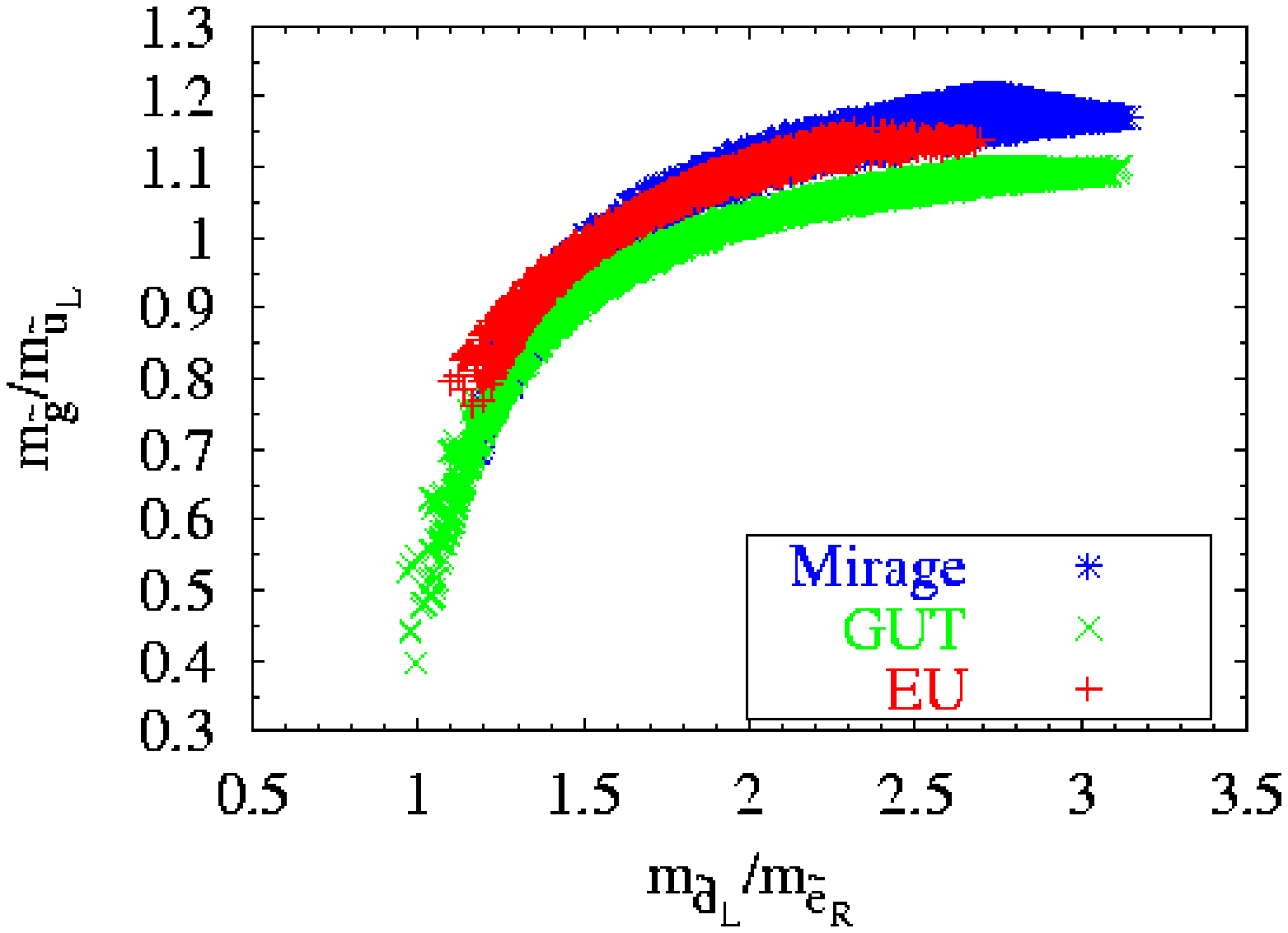, width=2.9in}}
\put(3,0){\epsfig{file=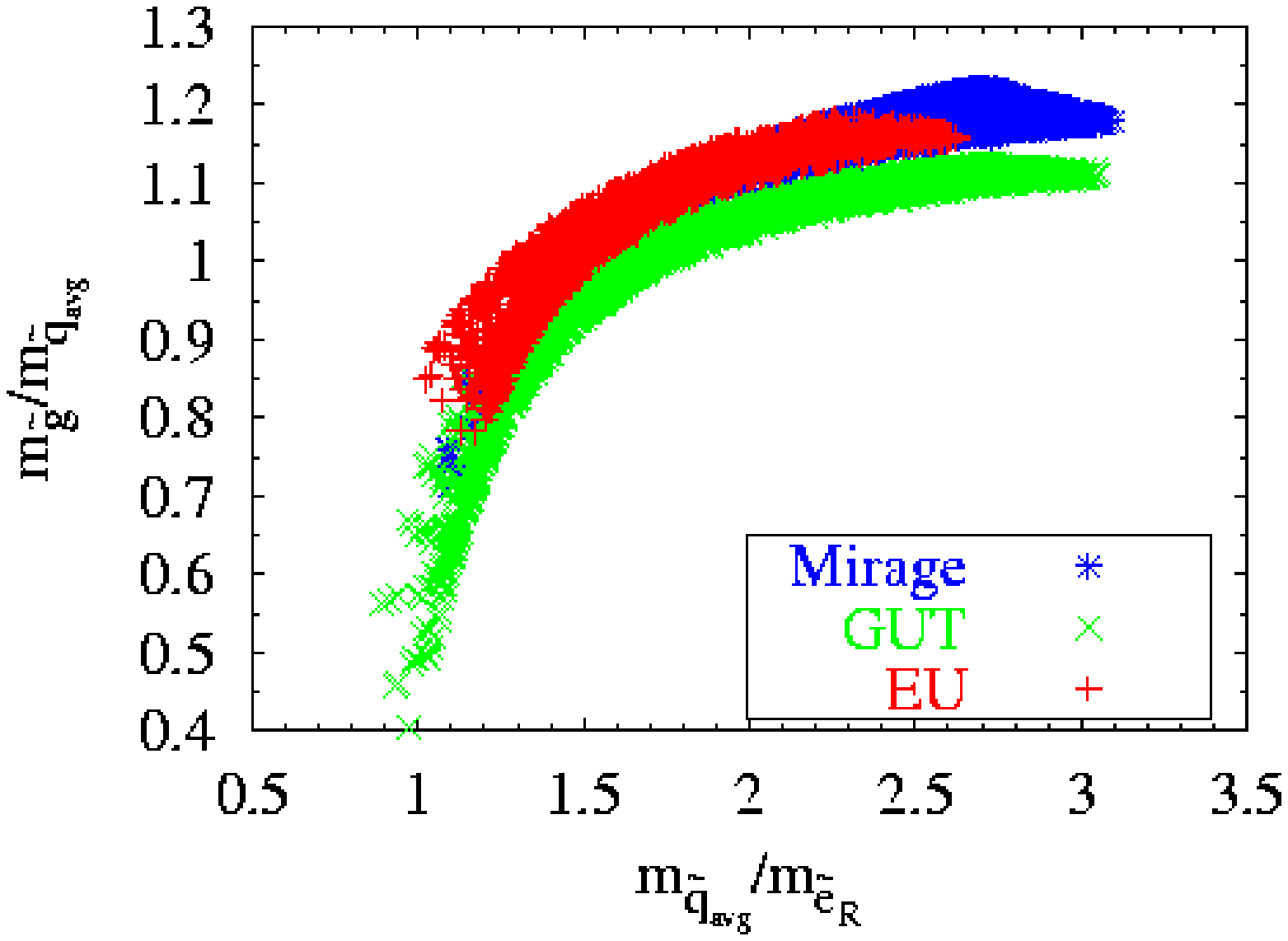, width=2.9in}}
\put(0.1,2.2){(a)}
\put(3,2.2){(b)}
\end{picture}
\caption{Highly discriminating ratios of sparticle masses. Each
plotted point corresponds to one data point in four-dimensional
parameter space ($\theta$, $\phi$, $\tan\beta$, $\mgr$) that is not
excluded by the imposed limits. In (b), $m_{\tilde{q}_{avg}}$ is as defined in
the text.
Pluses correspond to the early unification scenario, crosses to GUT-scale
unification and stars to mirage unification. 
}
}
Each of the three scenarios
were scanned over and shown on the plots. 
We have separated figure~\ref{allscat} into two cases: either the squark
flavours/handedness are known, as assumed in figure~\ref{allscat}a, or a
squark mass averaged over the first two generations and handedness is used, 
\begin{equation}
m_{{\tilde q}_{avg}} \equiv \frac{1}{8} \left( m_{{\tilde u}_L} + m_{{\tilde
u}_R} 
+ m_{{\tilde c}_L} + m_{{\tilde c}_R} + m_{{\tilde d}_L} + m_{{\tilde d}_R} +
m_{{\tilde s}_L} + m_{{\tilde s}_R}\right),
\end{equation}
as in
figure~\ref{allscat}b. For our numerical analysis we assumed degeneracy between
the first two families, and therefore 
in all our figures, $m_{{\tilde u}_L}$ and $m_{{\tilde d}_L}$
can also stand for $m_{{\tilde c}_L}$ and $m_{{\tilde s}_L}$ respectively. 
The reason for showing both figures is the differing ability of hadron
and lepton collider machines. It will be impossible for a hadron machine like
the LHC to distinguish
between the flavour of the almost-degenerate first two families of squarks. 
It is also difficult to see how a mass measurement of different handedness of
squarks might occur. 
The third family typically are split from these in mass, and have additional
heavy flavour tagging identification for their decay products and so it should
be possible to separate these. 
On the other hand, possible future linear $e^+ e^-$ colliders are
expected~\cite{linear} to provide very precise measurements of $m_{{\tilde
e}_L}$
and $m_{{\tilde e}_R}$. Because a future linear collider could scan to an
energy that is sufficient to
produce a particular mass of squark, it will be possible to separate up-type
from down-type squarks provided there is enough energy to produce them, and
identify which type squarks are produced from their cross-section. Handedness
could either be determined by examining the decay chains, or by using
polarised beams to favour production of one handedness or another.
It is conceivable that squarks will only be produced at the LHC, and are too
heavy for a particular linear collider facility to produce. It is for this
reason that we provide discrimination plots for both individual and averaged
squark masses. The importance of possessing individual squark mass
measurements can then be assessed. 

In figure~\ref{allscat}a there  
is a large overlap between the early unification and mirage scenarios'
predictions in the 
range $0.75<m_{{\tilde g}}/m_{{\tilde u}_L}<1.15$, but outside of this range
separation is possible. GUT-scale unification can usually be discriminated from
the 
other two scenario ranges, provided we measure the mass ratios with
$1\%$ accuracy.
In
Fig.~\ref{allscat}b, we see that the effect of averaging over the squark mass
measurements is to significantly decrease the discrimination between the
mirage and GUT unification scenarios.
There is significant overlap in the allowed
ranges for all three scenarios, but that there are also regions of predicted
mass ratios which can are only consistent with one particular scenario towards
the bottom-left or top-right of the plot.
If the empirical uncertainties on mass ratios are small enough and the
underlying model does not lie in an overlap region, the correct
scenario would be selected over the other two. 

It is possible that other measurements (for example measuring Higgs couplings)
will constrain some of the parameters of the model. 
For this reason, we will now assume that two of the four variables
$\theta$, $\phi$, $\tan\beta$ or $\mgr$ are constrained in order to see the
resulting improvement in string scenario discrimination. 
An obvious example to examine is the dilaton domination scenario
(corresponding to $\theta=90\dg$), which has attracted attention in the past in
its own right.
In figure~\ref{scat-dil}a, we 
plot the ratios of masses $x\equiv m_{\tilde e_L}/m_{\tilde e_R}$
  against $y\equiv m_{\tilde g}/m_{\tilde{u_L}}$ 
in the dilaton domination limit for all scanned values of
the gravitino mass and $\tan\beta$ in table~\ref{input}. We can
 clearly distinguish between the three scenarios because there is no overlap
in their regions in ratio parameter space. 
From the figure, we note that 
at least a\footnote{We always quote the fractional uncertainty, \eg on a 
ratio $R$, we quote the error required upon $\Delta R/R$.} 3$\%$ measurement
of $m_{\tilde g}/m_{\tilde{u_L}}$ and a $4\%$
measurement in $m_{\tilde e_L}/m_{\tilde e_R}$ would be required to 
always separate
the models on the basis of these ratios alone. 
This is a very positive result: assuming
dilaton domination, 
we have complete discrimination. Figure~\ref{scat-dil}b shows that there are
minimal changes if one averages over the first two generations of squarks. The
discriminating accuracy becomes smaller: 1$\%$ and 3$\%$ in 
$m_{\tilde g}/m_{{\tilde q}_{avg}}$ and $m_{\tilde e_L}/m_{\tilde e_R}$
respectively. 
\FIGURE[t]{\twographs{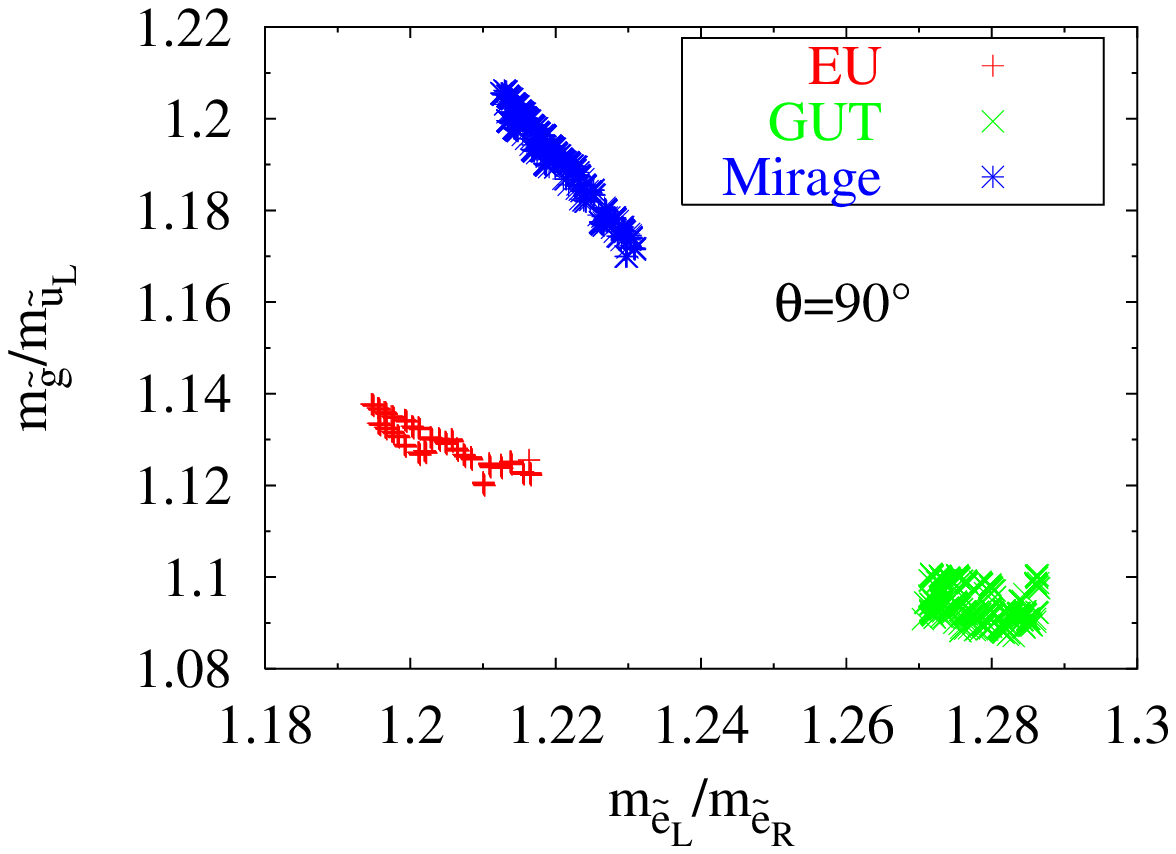}{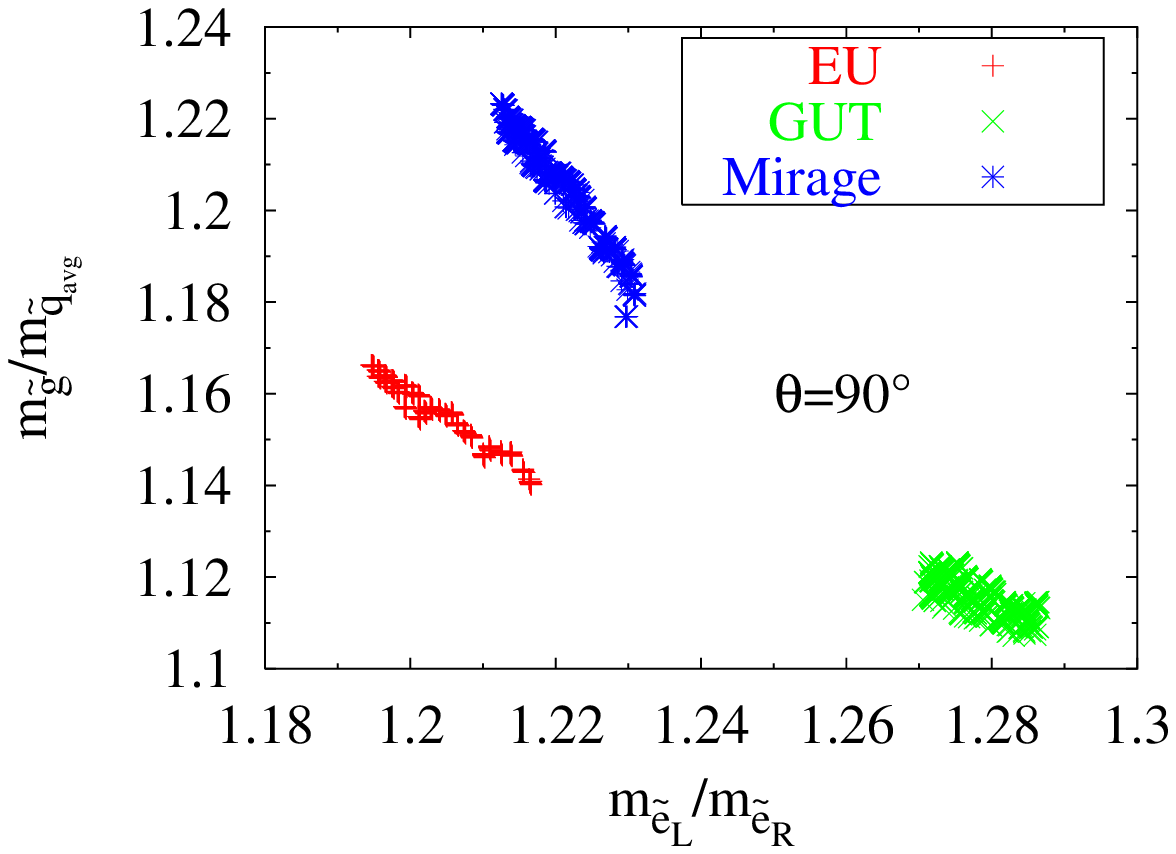}
\caption{\label{scat-dil}
Highly discriminating ratios of sparticle masses for dilaton domination
($\theta=90\dg$). 
Each plotted
point corresponds to one $(\mgr,\tan\beta)$-pair in the scan not excluded by 
any limits. Green crosses are predicted by the GUT unification scale, early
unification is denoted by red crosses, and blue stars are valid for the
mirage unification scenario.}
}

Figure~\ref{BMscat} illustrates departures from dilaton domination. 
$m_{3/2}$ and $\tan\beta$ have been fixed in the figure, whereas
$\theta$ and $\phi$ have been scanned over as in table~\ref{input}.
\FIGURE[t]{%
\label{BMscat}
\fourgraphs{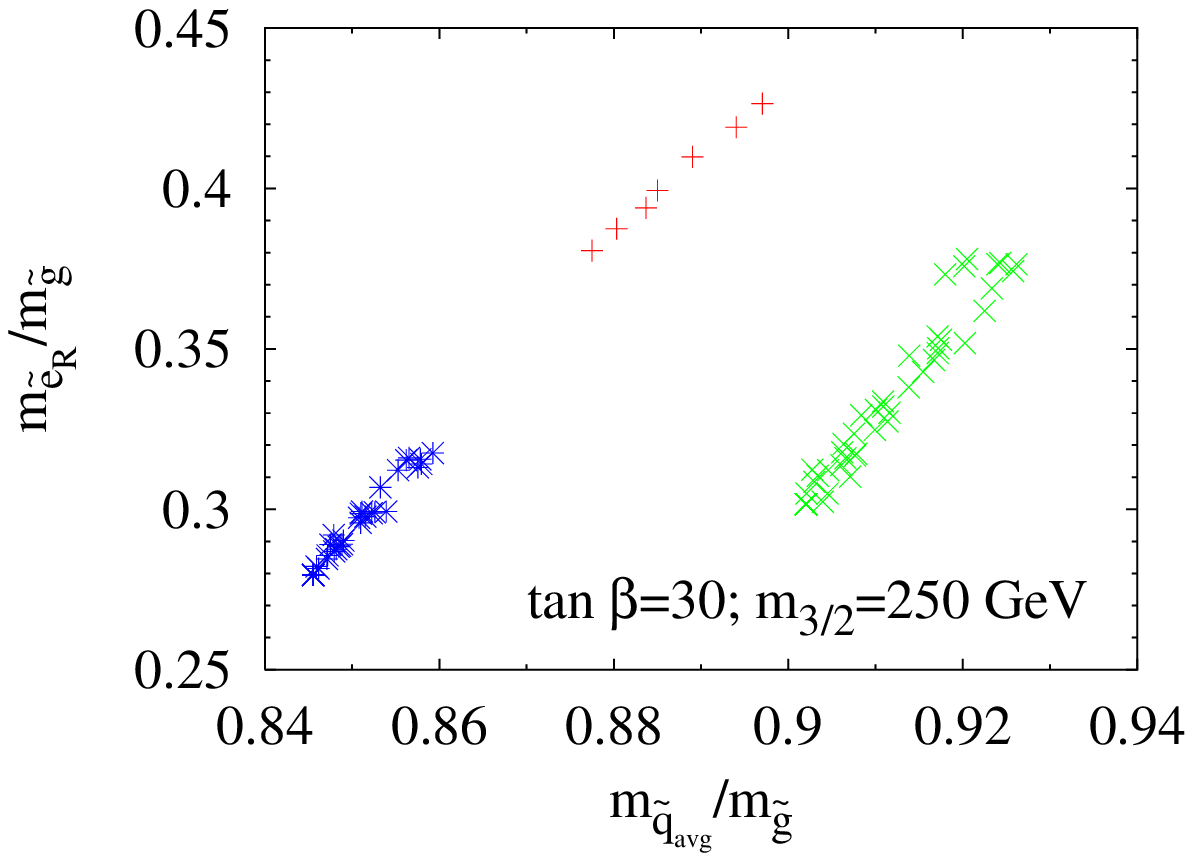}{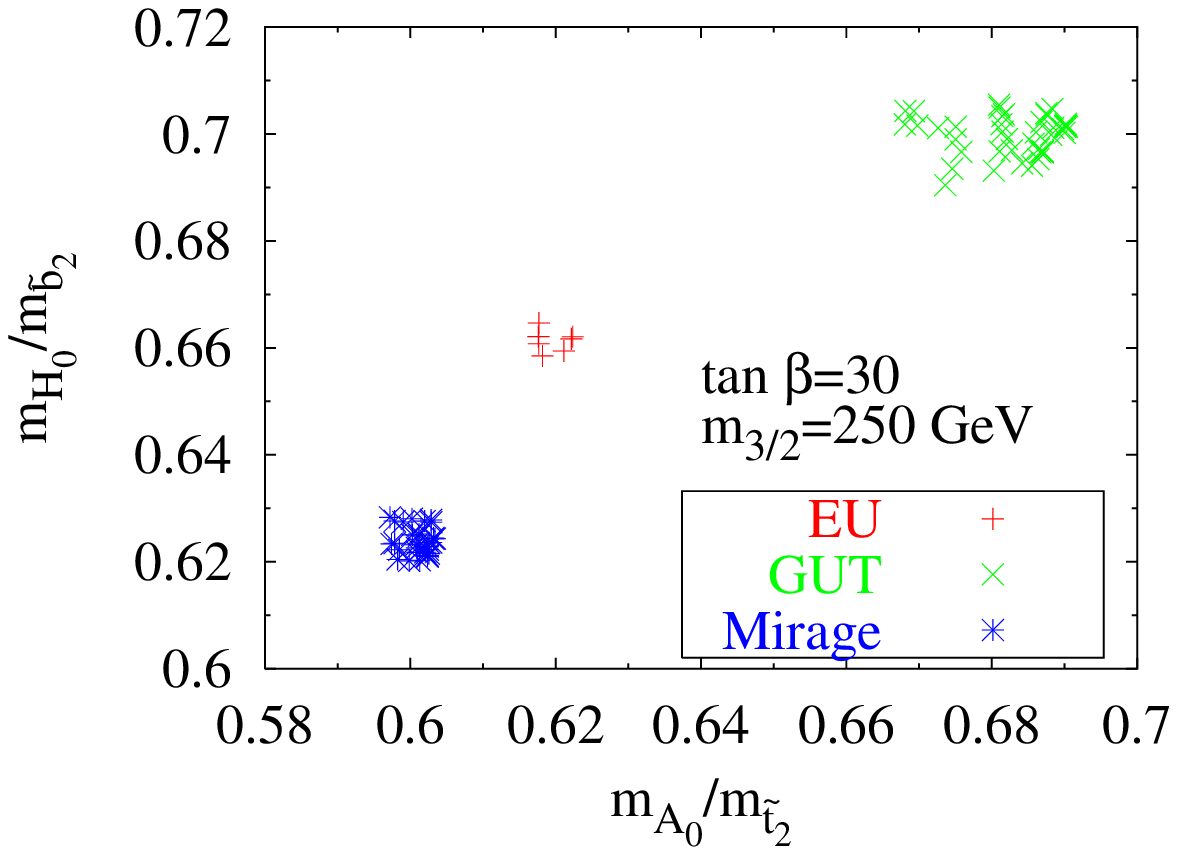}{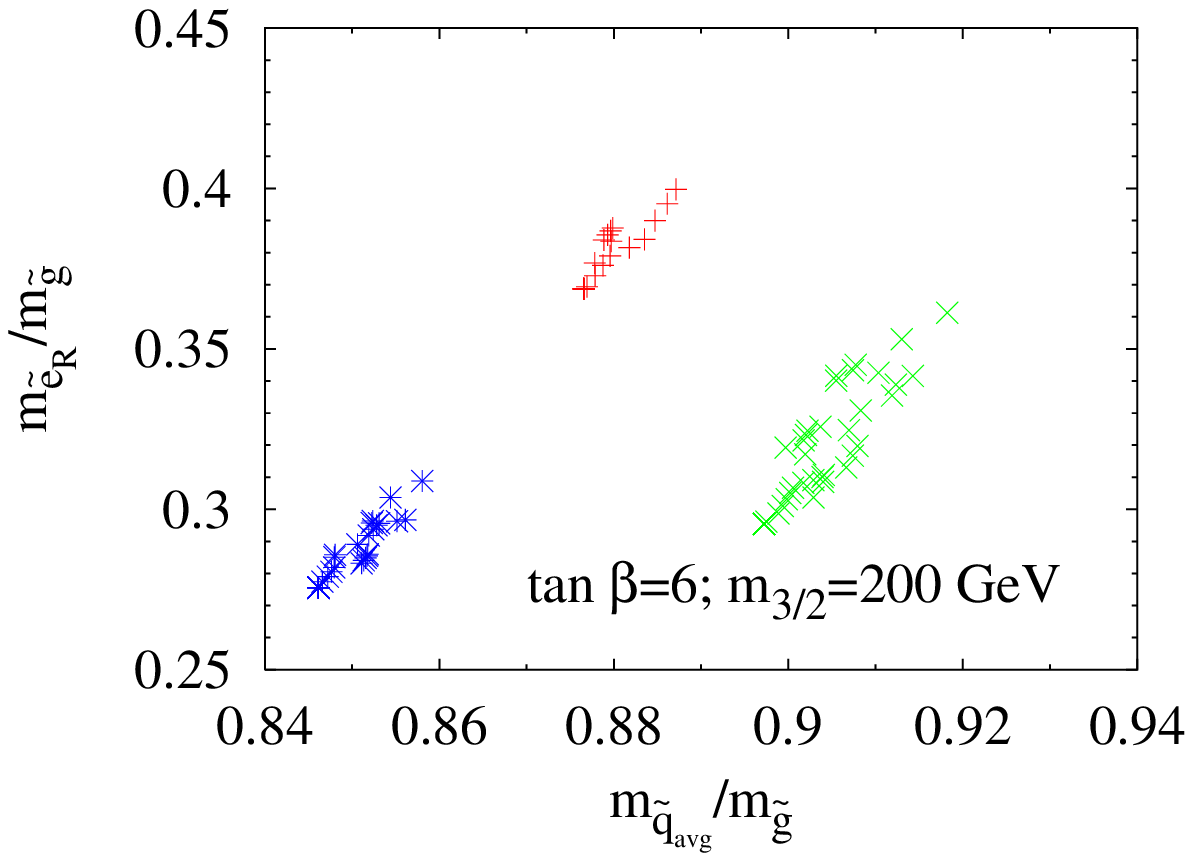}{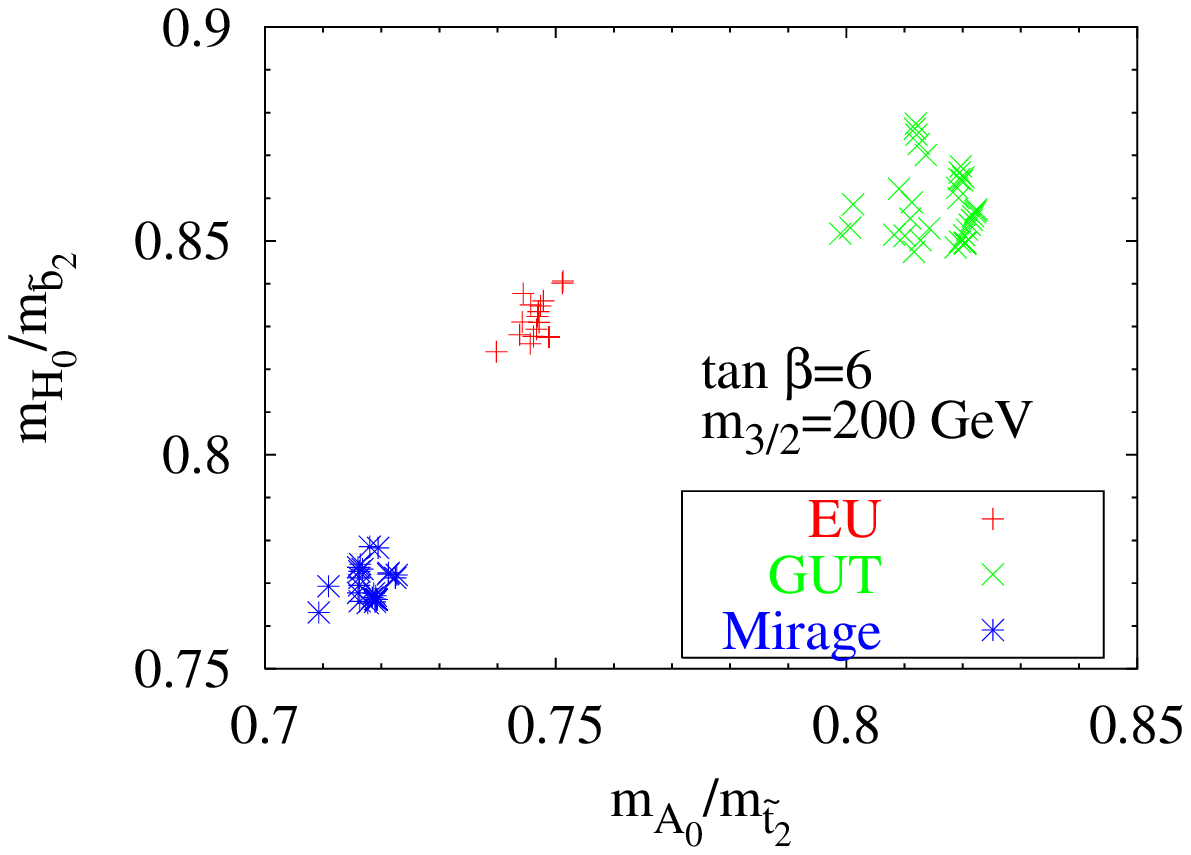}
\caption{Highly discriminating
ratios of sparticle masses for (a), (b): $\tan\beta=30$,
$\mgr=250\gev$ and (c), (d): $\tan\beta=6$,
$\mgr=200\gev$. Each plotted
point corresponds to one $(\theta,\phi)$-pair in the scan not excluded by 
the bounds. Green crosses are predicted by the GUT unification scale, early
unification is denoted by red crosses, and blue stars are valid for the
mirage unification scenario. 
}}
The figure shows the variation of the ratios $m_{\tilde{e}_R}/m_{\tilde{g}}$
and $m_{{\tilde q}_{avg}} / m_{\tilde{g}}$ as well as the variation of
$m_{H^0}/m_{\tilde{b}_2}$ 
and $m_{A^0} / m_{\tilde{t}_2}$ with string scenario. 
Figures~\ref{BMscat}a and~\ref{BMscat}b  
have $\tan \beta=30$ and $m_{3/2}=250$ GeV fixed to different values than
figures~\ref{BMscat}c and~\ref{BMscat}d ($\tan \beta=6$ and $m_{3/2}=200$ GeV),
in order to
illustrate how the required discrimination accuracy can depend upon the
supposedly constrained parameters.
Discrimination is shown to generally be achieved for errors smaller than 14$\%$
in $m_{{\tilde e}_R}/m_{\tilde g}$ and 4$\%$ in $m_{{\tilde
q}_{avg}}/m_{\tilde g}$ from figure~\ref{BMscat}a. 
These numbers happen to be roughly identical for
a different 
choice of $\tan \beta$ and $m_{3/2}$, as shown in figure~\ref{BMscat}c. 
Figures~\ref{BMscat}b,d show how discrimination could
be achieved with 5,7$\%$ errors respectively on $m_{H^0}/m_{\tilde{b}_2}$ and 
$m_{A^0}/m_{\tilde{t}_2}$. We note that the overall values of these two
ratios in any one scenario depends sensitively upon the values of 
$\tan \beta$ and $m_{3/2}$. $\tan \beta$ and $m_{3/2}$ could then be
constrained by their measurement.

\FIGURE[t]{%
\twographs{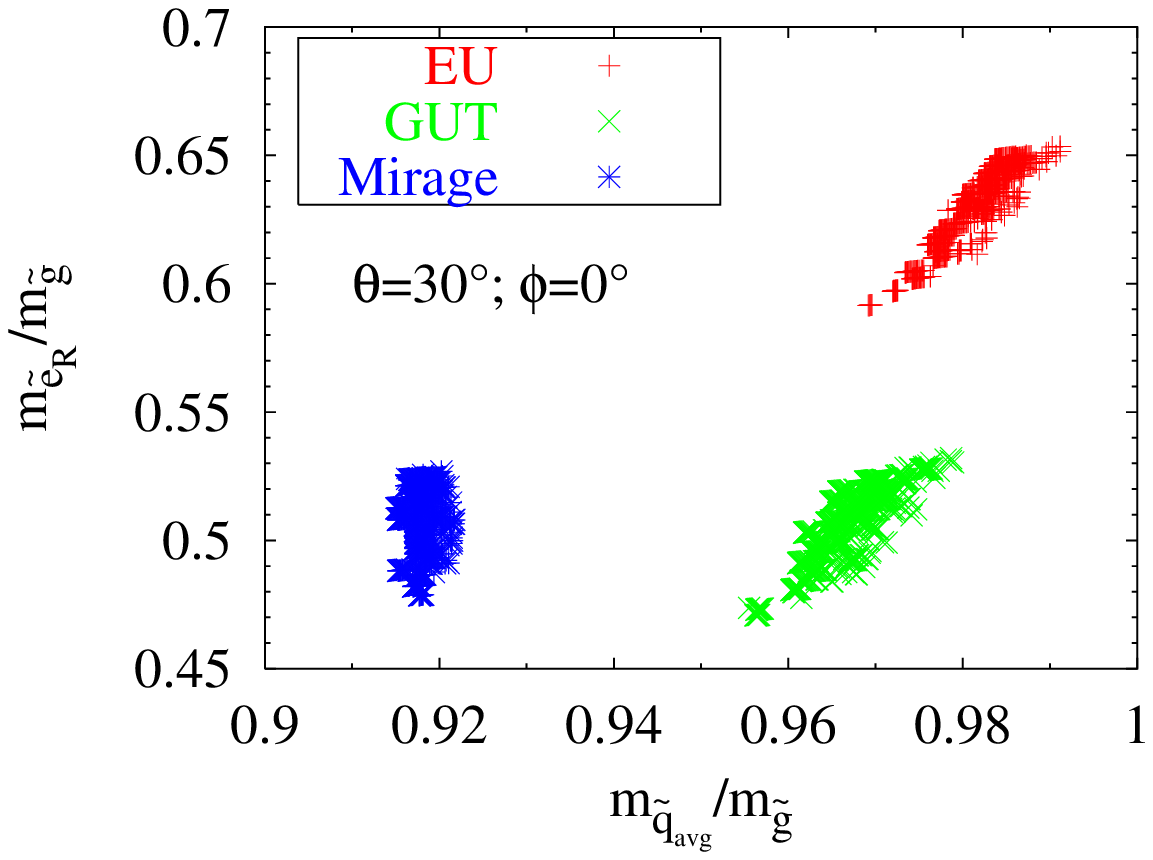}{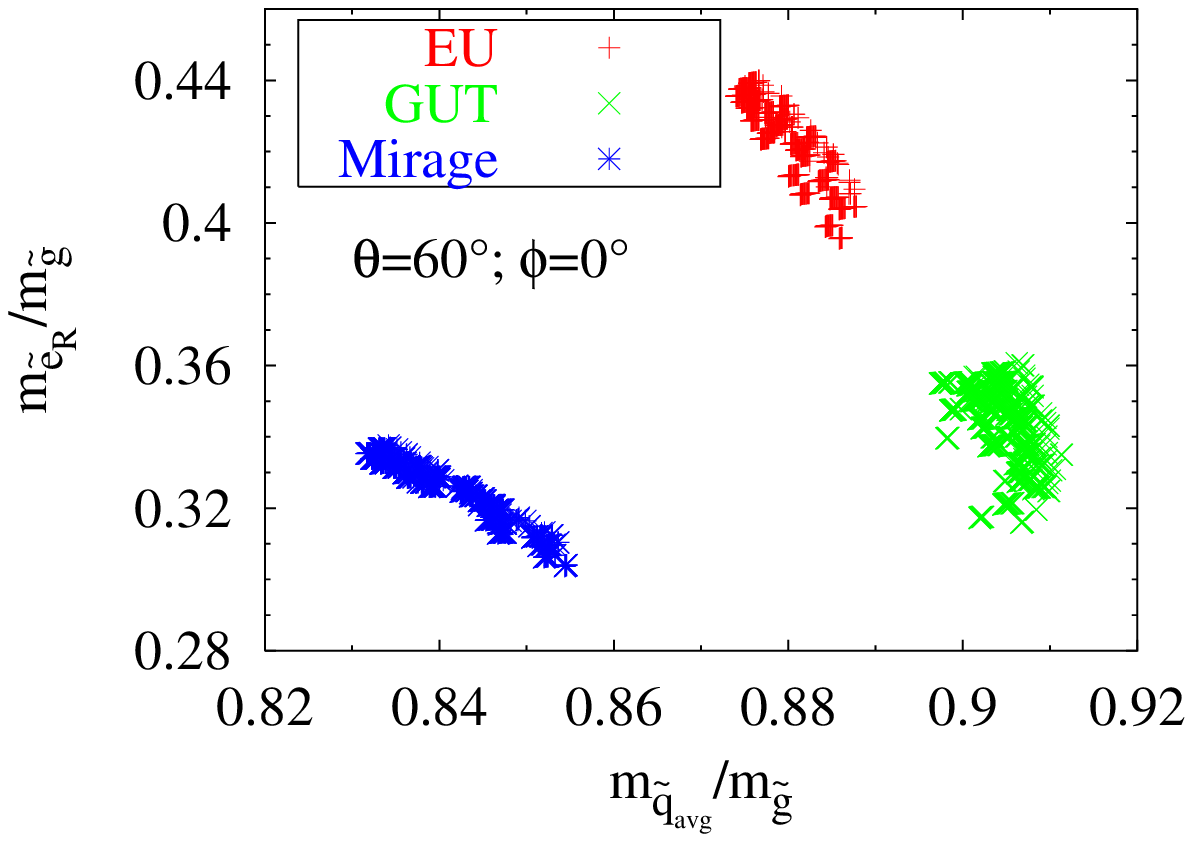}
\label{together}
\caption{%
Discriminating ratios of sparticle masses in departures from dilaton
domination: (a) $\theta=30\dg, \phi=0\dg$ and (b) $\theta=60\dg$,
$\phi=0\dg$. Each plotted point corresponds to one $(m_{3/2},\tan\beta)$
pair in the scan not excluded by the limits.
Green crosses are predicted by the GUT unification scale, early
unification is denoted by red crosses, and blue stars are valid for the
mirage unification scenario. 
}}
We now fix the goldstino angles, to see if it is possible to distinguish the
models when they are scanned over $m_{3/2}$ and $\tan \beta$. 
The usefulness of combinations of ratios of sparticle masses depends in
general upon the value of these fixed
goldstino angles. As figures~\ref{together}a-\ref{together}b show, however, 
that the required accuracy on the ratios $m_{{\tilde q}_{avg}}/m_{\tilde{g}}$
and
$m_{\tilde{e}_R}/m_{\tilde{g}}$ does vary with respect to a
variation of the goldstino angles.
From figures~\ref{together}a,b we see that for $\phi= 0\dg$,
$(\theta=30\dg,\theta=60\dg)$ we require a $(3\%,5\%)$ 
measurement of $m_{{\tilde q}_{avg}}/m_{\tilde{g}}$ 
and a $(11\%,10\%)$ measurement of $m_{\tilde{e}_R}/m_{\tilde{g}}$ to
discriminate the three scenarios. In fact, these ratios discriminate 
for 
any fixed value of $\theta, \phi$ but do not completely discriminate once the
goldstino angles are scanned over. 
It is therefore difficult to say much that is definitive about the ratios in
this case. It is possible that specific models will predict the goldstino
angles, and in that case the two ratios mentioned above will be useful.
However, the connection between measured quantities and $\theta, \phi$ is 
perhaps less obvious than for $m_{3/2}$ and $\tan \beta$.
This means that it is difficult to see how one would directly infer their
values from experimental data, and so we leave the
discussion of constrained goldstino angles here.

\subsection{Interpretation}
Note that the required accuracy quoted here is the one necessary
to separate the closest two points between different scenarios, \ie assuming
that the data lie within one of the predicted regions, what accuracy would 
definitely exclude the other two regions.
This is the
most 
pessimistic scenario, and we may well require less accuracy depending upon
where the data finally lie. For example figure~\ref{allscat}a shows that the data
$m_{{\tilde d}_L}/m_{{\tilde e}_R} \sim 3.1 \pm 0.6$ 
and $m_{\tilde g}/m_{{\tilde u}_L} \sim 1.05 \pm 0.03$, 
would imply that the GUT-scale scenario would be selected over the
other two whereas for 
$m_{{\tilde d}_L}/m_{{\tilde e}_R} \sim 1.8$ 
and $m_{\tilde g}/m_{{\tilde u}_L} \sim 1.0$,
uncertainties of (0.2,0.02) would be required respectively.
Where the data finally lie will also decide the most important measurement.
We take the dilaton dominated scenario to exemplify this point
(see figure~\ref{scat-dil}a). If the data lie over the GUT-scale region, the
most 
important errors are those upon $m_{{\tilde e}_L}/m_{{\tilde e}_R}$ as this
will best separate the GUT-scale scenario from the other two. If, on the other
hand, the data lie over the early unification scenario, it is more important to
have smaller uncertainties upon $m_{\tilde g}/m_{{\tilde u}_L}$ for
discrimination against the mirage unification scenario.

We must ask what effect theoretical errors have on the interpretation of the
results. Comparisons between the spectra derived from 
{\tt SOFTSUSY}~\cite{softsusy}
code and {\tt ISASUGRA}~\cite{isajet}, 
while showing qualitative agreement, display
differences of order of the mass differences required to discriminate the
scenarios~\cite{snowmass,softsusy}. The errors are both dependent upon
the flavour of particle (sleptons tend to have very good agreement, 
whereas coloured objects typically 
show a 3-5$\%$ difference) and on the region of parameter space. For extreme
places in parameter space, for example near the radiative electroweak symmetry
breaking boundary, predicted mass differences can be large between the two
codes~\cite{snowmass,softsusy}. However, provided the point in parameter space
being 
examined is not particularly special, any theoretical errors in codes should
not change our 
conclusions about discrimination. This is because the errors due to (for
example) not including finite threshold corrections in {\tt ISASUGRA} will affect
all points in the three scenarios in roughly the same way, moving them all in
one direction for example. The statistical errors $m_t=175\pm5$ GeV similarly
provide an uncertainty. We note that this error should be drastically reduced
by Run II of the Tevatron.

It is clear, however, that better accuracy in the mass predictions will be
necessary when interpreting real data.
Of course, the discrimination depends at face value on the assumption
of prior knowledge either of the goldstino angles (which will probably not be
owned) or of $m_{3/2}$ and $\tan \beta$. The latter pair of parameters may
be predicted in the hypothesis of each scenario, since 
gaugino masses can help us divine
$m_{3/2}$ and $\tan \beta$ may well be obtained
from Higgs measurements~\cite{linear}. 
We note that each discriminating plot is valid for two particular values of
fixed parameters.
If the fixed parameters are not determined beforehand, they would have to be
fit to the sparticle masses.
This fitting procedure will be subject more to the theoretical errors, and will
require them to be shrunk from their current levels. 
We would of course use all available (\ie measured) highly discriminating
ratios simultaneously in order to try to pin down the correct scenario. 
This would provide additional confirmation that the identified scenario is the
correct one.

It is likely that the next machine capable of producing TeV-scale
superparticles in sufficient numbers to make a reasonable measurement is the
LHC\@. It remains to be seen how good the LHC is in determining ratios of
masses, but it
cannot constrain any single sparticle mass to a very accurate
level (10$\%$ or so on the absolute values).
The LHC can, however, constrain certain
functions of masses to about 
1\%~\cite{yao,atlasCMS,ATLASTDR,us} by finding the end-points of kinematic
distributions. 
In ref.~\cite{us}, the three most accurate measurements were
found to be on the edges in the invariant masses detailed in table~\ref{edges}.
They follow parts of the decay chain of 
${\tilde q} \rightarrow q \chi_2^0$, $\chi_2^0 \rightarrow l {\tilde l}$, then 
${\tilde l} \rightarrow l \chi_1^0$. The existence of this chain depends upon
the mass ordering $m_{\tilde q} > m_{\chi_2^0} > m_{\tilde l}$.

The edges shown in table~\ref{edges} measure~\cite{us}
\begin{eqnarray}
(m_{ll}^{\mbox{\tiny max}})^2 &=& \frac{(m_{\chi_2^0}^2 - m_{{\tilde l}_R}^2)
(m_{{\tilde l}_R}^2 - m_{\chi_1^0}^2)}{m_{{\tilde l}_R}^2}, \nonumber \\
(m_{lq \mbox{\tiny (low)}})^2 &=& \mbox{min} \left[ 
\frac{(m_{\tilde q}^2 - m_{\chi_2^0}^2)
(m_{\chi_2^0}^2 - m_{{\tilde l}_R}^2)}{m_{\chi_2^0}^2}, \ 
\frac{(m_{\tilde q}^2 - m_{\chi_2^0}^2) (m_{{\tilde l}_R}^2 -
m_{\chi_1^0}^2)}{2 m_{{\tilde l}_R}^2 - m_{\chi_1^0}^2}
\right], \nonumber \\
(m_{llq}^{\mbox{\tiny max}})^2 &=& \mbox{max} \left[
\frac{(m_{\tilde q}^2 - m_{\chi_2^0}^2)(m_{\chi_2^0}^2 - m_{\chi_1^0}^2)}{m_{\chi_2^0}^2},\
\frac{(m_{\tilde q}^2 -m_{{\tilde l}_R}^2)(m_{{\tilde l}_R}^2
-m_{\chi_1^0}^2)}{m_{{\tilde l}_R}^2},\right. \nonumber \\
&& \left.
\frac{(m_{{\tilde l}_R}^2m_{\tilde q}^2 -m_{\chi_1^0}^2m_{\chi_2^0}^2)
(m_{\chi_2^0}^2 - m_{{\tilde l}_R}^2) }{m_{{\tilde l}_R}^2m_{\chi_2^0}^2}
\right], \label{Rs}
\end{eqnarray}
except for the special case in which $m_{{\tilde l}_R}^4 <m_{\tilde
q}^2m_{\chi_1^0}^2 < m_{\chi_2^0}^4 $ and
$m_{\chi_2^0}^4 m_{\chi_1^0}^2 < m_{\tilde q}^2m_{{\tilde l}_R}^4$ when
$(m_{llq}^{\mbox{\tiny max}})^2=(m_{\tilde q}-m_{\chi_1^0})^2$ instead.
We use $m_{ll}^{\mbox{\tiny max}}$ to normalise the other edge variables, as
its small error is negligible. 
\TABULAR[th]{|c|c|}
{\hline \label{edges}
edge variable & accuracy \\ \hline
$m_{ll}^{\mbox{\tiny max}}$ & 0.1$\%$ \\
$m_{llq}^{\mbox{\tiny max}}$ & 1$\%$ \\
$m_{lq \mbox{\tiny (low)}}$ & 1$\%$ \\
\hline
}{Accuracy of most useful LHC edge variables.}
\EPSFIGURE[t]{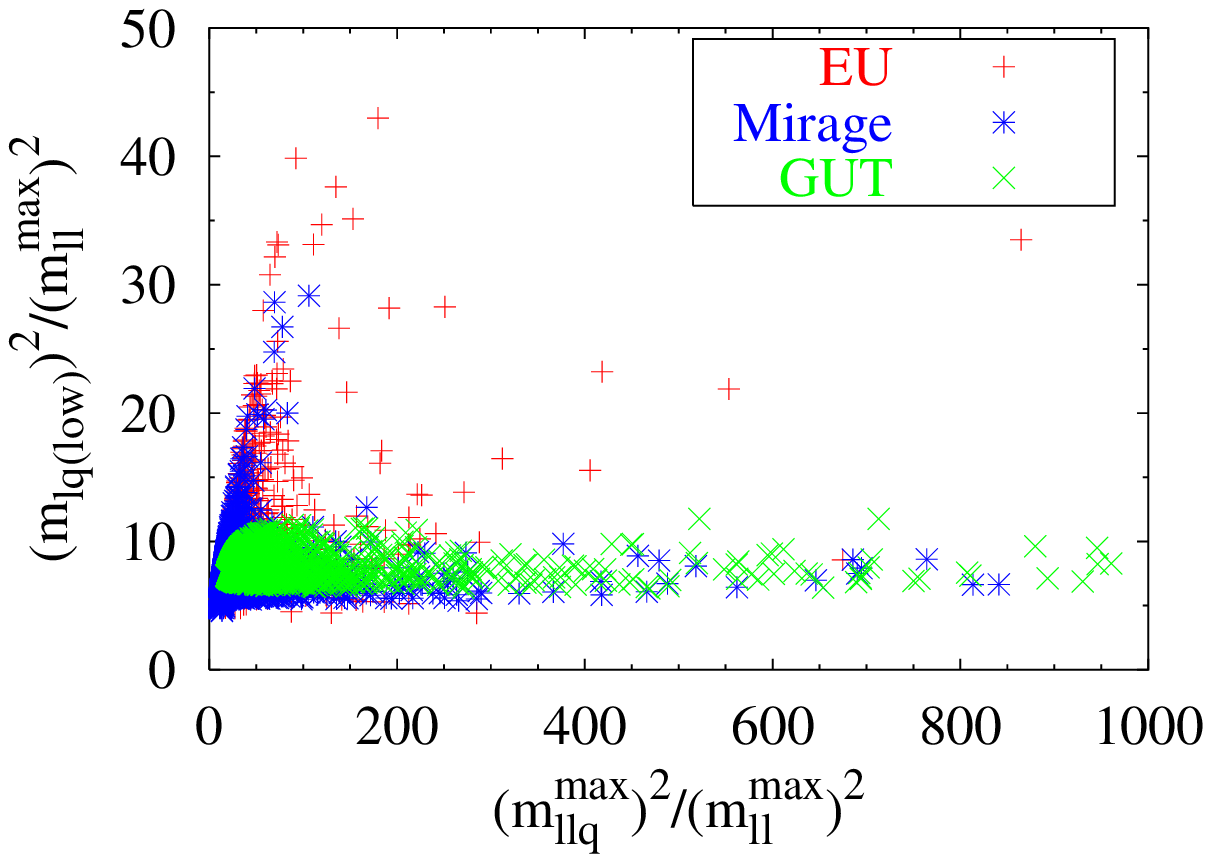}{\label{edge}
Ratios of LHC edge variables.
Each
plotted point corresponds to one data point in four-dimensional
parameter space ($\theta$, $\phi$, $\tan\beta$, $\mgr$) that is not
excluded by the imposed limits and where $m_{\chi_2^0}>m_{{\tilde l}_R}$.
}
We therefore plot the ratios $(m_{llq}^{\mbox{\tiny
max}})^2/(m_{ll}^{\mbox{\tiny max}})^2$ against $(m_{lq \mbox{\tiny
(low)}})^2/(m_{ll}^{\mbox{\tiny max}})^2$ in figure~\ref{edge} in order to see 
if these precision LHC measurements can be used to distinguish the string
scenarios. All four free parameters are scanned over and all current
experimental 
constraints (including $g-2$ of the muon) are applied. Additionally,
we require $m_{\chi_2^0}>m_{{\tilde l}_R}$, to make sure that the
above mentioned decay chain actually exists. This is the case in
roughly one third of the investigated scan points. 

As figure~\ref{edge} shows, there is no clear separation between all
scenarios on the basis of these LHC edge variables.
However, the $(m_{lq \mbox{\tiny (low)}})^2/
(m_{ll}^{\mbox{\tiny max}})^2$-axis provides some exclusion limits: 
the
GUT scale case could be ruled out if $(m_{lq \mbox{\tiny (low)}})^2/
(m_{ll}^{\mbox{\tiny max}})^2 $ does not lie between 7 and 12. 
The horizontal axis appears to contain no discriminatory power.

\section{Conclusions}
We have seen that general information can be obtained about 
 string theory scenarios. We considered three general string scenarios: 
two with the MSSM superfield content (GUT scale unification and mirage
unification at the intermediate scale) and one with additional leptons and
unification at the intermediate scale.
We have mapped out the available parameter space for each scenario, using
current empirical constraints. 

Our main results are better  summarised in figures~\ref{maps1} and~\ref{allscat}. 
Figure~\ref{maps1} provides a map of viable parameter space slices
for each of the three scenarios.
We can see clearly that for different combinations of values of the
goldstino angles different
regions of $m_{3/2}-\tan\beta$ parameter
space are allowed.

In figure~\ref{allscat}
we find that ratios of particle masses 
 may be able to separate each
scenario when {\it all} \/parameters: $\theta,\phi,m_{3/2},\tan\beta$
are considered. This allows us to differentiate between the three
scenarios provided the data do not appear in certain overlap regions. Clearly,
fixing some of 
the parameters makes the separation between scenarios cleaner (see
figs.~\ref{scat-dil}-\ref{together}).
 We then determined what accuracy is required
on the measurement of these masses in order to distinguish the scenarios. 
The errors are required to be less than the few percent level, depending upon
the exact ratio.
 
A previous study~\cite{lepts} found that the exotic heavy leptons present in 
the early unification scenario, can be discovered by the LHC if their masses
are less than 980 GeV. This could confirm the mirage scenario, making further
discrimination redundant between the models considered here. 
The exotic leptons would 
not be discovered if they were heavier than 1 TeV, so all three scenarios
would still be left to be distinguished in that case.
We note that a previous study 
confronted a fundamental SUGRA model with
detailed linear collider and LHC measurements in a bottom-up
approach~\cite{grah}. Presumably, this kind of analysis could be repeated for
stringy scenarios and would complement the
present study. It could be used to provide accurate measurements of high-scale
soft-breaking parameters and to determine at what scale they unify.

The small percent-level errors required on mass ratios are not expected to
occur at the Tevatron experiments~\cite{tevat}, or even at the
LHC~\cite{ATLASTDR}. But it is likely that a future linear $e^+ e^-$ collider
facility~\cite{linear} would have sufficient accuracy, provided it had
sufficient centre of mass energy to produce some of the particles involved in
the ratios. 
\TABULAR[t]{|c|c|c|c|c|c|c|c|c|c|}{\hline
& $\tilde{g}$ & $m_{\tilde{e}_R}$ & $m_{\tilde{e}_L}$ & $m_{\tilde{u}_L}$ & 
$ m_{\tilde{t}_1}$ & $ m_{\tilde{t}_2}$ & $M_{\tilde{\chi}_1^0}$  & $ m_{H^0,A^0}$\\ \hline
Early  & 380 & 180 & 230  & 410 & 190 & 450  & 100   & 170 \\
Mirage & 310 & 170 & 210  & 430 & 130 & 450  &  77   & 160  \\
GUT    & 320 & 160 & 200  & 450 & 180 & 470  &  55   & 190 \\ \hline}
{\label{rangesH} Lower bounds on relevant sparticle masses coming from the parameter scans, in GeV.
The rows are marked by string scenario: early unification, mirage and GUT-scale unification
respectively.}
For this reason, we present the lower bounds on relevant sparticle masses in
table~\ref{rangesH} for each scenario. Notice that this is an explicit
way to differentiate the different scenarios. If, for instance, the
neutralinos are discovered with a mass smaller than $60$ GeV this
would clearly favour the GUT scenario over the other two.
Although the LHC can cover the lower bounds on each sparticle mass,
only a multi TeV facility such as CLIC
would provide enough centre of mass energy to produce sparticles if their
masses are several TeV.
We note, however, that lower values of $m_{3/2}$ (and therefore lower sparticle
masses) are favoured by the
fine-tuning parameter~\cite{aaikq}.

There are several directions in which our work can be extended. First 
we may incorporate the charge and colour breaking (CCB) constraints on all 
the scenarios, as was done for the dilaton domination scenario in
ref.~\cite{aaikq}. While it would certainly be interesting to 
know if the universe were in some meta-stable vacuum, a global CCB vacuum
might not necessarily rule the model out~\cite{ccbno}. 
We could impose some arbitrary fine-tuning 
constraint on the models in order to restrict the maximum values of $m_{3/2}$;
this would presumably increase the discriminatory power of our mass ratio
tests. 
Another extension relates to the consideration of moduli 
domination, a limit that we did not consider since in that case the
soft breaking terms appear only at the loop level and therefore
anomaly mediation cannot
 be neglected. It is an open question to correctly combine anomaly mediation 
with gravity mediation for all soft breaking terms. 

This study is a crude first step towards the experimental discrimination of 
string models, leaving plenty of room for more investigation. 
For example, we only considered two mass ratios at a time for
discrimination, but this could be generalised to higher numbers, requiring an
automated algorithm for finding the most discriminating ratios.
It would be interesting to know the search reach in parameter
space of the various future colliders as could be performed by
{\tt HERWIG}~\cite{herwig}. 
Further, we would like to know 
under which conditions experiments will be able to measure the discriminatory
mass ratios. 

We hope that the techniques introduced in this article 
can be used to discriminate other SUSY breaking scenarios, assuming
eventual discovery of supersymmetric particles.

\acknowledgments
This work is funded by the Studienstiftung des deutschen Volkes 
 and the UK Particle Physics and Astronomy Research Council (PPARC).
 DG and FQ thank CERN Theory division for their hospitality.
DG thanks H Baer for advice on 
the {\tt ISASUGRA} package. BCA thanks J Ellis for
providing inspiration and we would all like to thank members of the Cambridge
LHC SUSY working group for interesting discussions on more experimental
aspects.

\end{document}